\documentclass[reprint,aps,pra]{revtex4-2}
\usepackage{amsmath}
\usepackage{graphicx}
\usepackage{amsfonts}
\usepackage{color}
\usepackage{hyperref}
\usepackage{cryptocode}
\usepackage{verbatim} 
\hypersetup{
	colorlinks = true,
	linkcolor =blue,
	citecolor=blue, 
	urlcolor=blue 
}
\usepackage{braket}
\usepackage{bm}
\newcommand{\mk}[1]{\textcolor{red}{#1}}

\graphicspath{{figures/}}

\usepackage{amsthm}
\newtheorem{remark}{Remark}

\newcommand{\Sfeas}{\mathcal{S}}
 
\begin{document}
	\title{Using Cascade in Quantum Key Distribution}  
	\author{Devashish Tupkary}
	\email{tupkary.devashish@gmail.com}
	\affiliation{Institute for Quantum Computing and Department of Physics and Astronomy, University of Waterloo, Waterloo, Ontario, Canada, N2L 3G1}	
	
	\author{Norbert L\"utkenhaus}
	\email{nlutkenhaus.office@uwaterloo.ca}
	\affiliation{Institute for Quantum Computing and Department of Physics and Astronomy, University of Waterloo, Waterloo, Ontario, Canada, N2L 3G1}

	\begin{abstract}
	We point out a critical flaw in the analysis of Quantum Key Distribution (QKD) protocols that employ the two-way error correction protocol Cascade. Specifically, this flaw stems from an incomplete consideration of all two-way communication that occurs during the Cascade protocol. We present a straightforward and elegant alternative approach that addresses this flaw and produces valid key rates. We exemplify our new approach by comparing its key rates with those generated using older, incorrect approaches, for Qubit BB84 and Decoy-State BB84 protocols. We show that in many practically relevant situations, our rectified approach produces the same key rate as older, incorrect approaches. However, in other scenarios, our approach produces valid key rates that are lower, highlighting the importance of properly accounting for all two-way communication during Cascade. 
	\end{abstract}

	\maketitle

\section{Introduction}
Quantum Key Distribution (QKD) \cite{bennett2014Quantum,bruss1998Optimal,bennett1992Quantum} can provide information-theoretic security of secret keys between two communicating parties, Alice and Bob. Since the quantum channel connecting Alice and Bob is not perfect in any practical realization, QKD protocols implement an error-correction step to correct errors in the measurement data collected by Alice and Bob. This involves classical communication between the two parties, and leaks additional information to the eavesdropper Eve, which must be accounted for when calculating the achievable secret key rate. Cascade \cite{brassard1994Secretkey} is one of the most widely used two-way error correction protocol for QKD. A lot of work has been done optimizing various parameters of the Cascade protocol, such as its blocksizes, number of rounds, efficiency etc  \cite{reis2019Quantum,pedersen2015High,martinez-mateo2015Demystifying,elkouss2009Efficient,calver2011empirical,mao2021High,pacher2015information,erven2008Entangled}. Cascade has also been used in a large number of QKD experiments \cite{dixon2017Quantum,su2009Continuous,gobby2004Quantum,tentrup2019Largealphabet,lorenz2004Continuousvariable}.
	
Our main result is to rectify a flaw in the analysis of QKD protocols using Cascade, which stems from an incomplete consideration of the two-way classical communication during Cascade. We observe that in past literature, only the communication from Alice to Bob has been accounted for when considering information leakage about the key to Eve. For a rigorous security proof, the communication from Bob to Alice must also be included when bounding the information leaked to Eve. 

We propose a straightforward and elegant alternative approach that produces valid key rates. The main idea is to compute key rates for a protocol that leaks all the communnication from Alice to Bob during Cascade, along with all location of errors in Alice and Bob's raw data to Eve, in the information reconciliation step. We show that this leaks more information to Eve than Cascade, and thus any key rate for such a protocol is a valid key rate for the original protocol that uses Cascade.

We apply our solution to the qubit based BB84 protocol, and the polarization encoded weak coherent pulse (WCP) BB84 with decoy intensities, for a variety of channel models and constraints.  We use the numerical framework from \cite{Winick2018} for our calculations. In this work, we restrict our attention to the asymptotic regime for simplicity, where one can assume an IID collective attack without loss of generality \cite{Renner2016, christandl2009Postselection}. However, our solution can be directly adapted to the analysis of finite size protocols. This is because many such analyses ultimately involve the optimization of the same objective function \cite{George2021, valerio_2008, valeriotwoway, Tomamichel_2017} (with different constraints), and our approach only modifies the objective function.
	 
This paper is organized as follows.  In Sec. \ref{sec:background} we explain the steps in a generic QKD protocol and explain the Cascade protocol briefly. In Sec. \ref{sec:usingcascade} we explain the problem with past analysis of QKD protocols using Cascade, and present our arguments for correcting it. We also review the numerical framework that we used to compute key rates in this work. In Sec. \ref{sec:qubit_bb84} and Sec. \ref{sec:WCP_Decoy} we apply our solution to the BB84 protocol implemented using qubits, and WCP states with decoy intensities respectively. In Sec. \ref{sec:conclusion} we present concluding remarks.

\section{Background}
\label{sec:background}
\subsection{Protocol Description}
In this subsection, we give a description of the asymptotic formulation of a typical QKD protocol that can use Cascade in the information reconciliation step. 
\begin{enumerate}
	\item \textbf{Quantum Phase:} In an entanglement-based protocol, Alice and Bob receive states from a source and perform measurements on them. In a prepare-and-measure protocol, Alice prepares and sends signals to Bob, who measures them. The security analysis of a prepare-and-measure scheme can be reduced to that of an entanglement-based scheme using the source replacement scheme \cite{curty2004Entanglement}.
	\item \textbf{Acceptance Test (Parameter Estimation):} Alice and Bob announce the measurements obtained, and signals sent, for a small fraction of signals. They then perform a test to decide whether to abort or continue the protocol. This step is modelled as Alice and Bob performing some measurements given by POVMs $\{ \Gamma_k\}$, obtaining expectation values $\{ \gamma_k \}$. The POVMs and expectations values depends on whether the protocol implements ``fine-graining'' or ``coarse-graining'' during the acceptance test \cite{wang2021Numerical}, and the exact nature of the coarse-graining.
	\item \textbf{Classical processing:} For the remaining signals, Alice and Bob perform some blockwise processing of data. This involves operations such as public announcements and sifting to remove unwanted signals. Alice then implements a key map that maps her local data and the information exchanged in the blockwise processing, to her raw key.  
	\item \textbf{Error correction and verification:} Alice and Bob implement error correction by exchanging classical information. Cascade can be used in this step. Alice and Bob then compare a randomly chosen hash of their raw keys for error verification, and abort the protocol if the hashes do not match.
	\item \textbf{Privacy Amplification :} Alice and Bob choose a common two-universal hash function and apply it to their raw keys to generate their final secret key. 
\end{enumerate}
If $\tilde{A}, \tilde{B}$ denote announcements made by Alice and Bob in the blockwise processing step, and $Z$ denotes the result of the key map implemented by Alice, and $E$ is Eve's quantum system, then the key rate is given by \cite{Renner2016,Devetak2005}
\begin{equation} \label{eq:basickey rate}
	R = \min_{\rho \in \Sfeas(\vec{\gamma})}	 S(Z| E \tilde{A} \tilde{B} ) - p_{\text{pass}} \times \delta_\text{leak},
\end{equation}
where the minimization is over all states $\rho$ belonging to $\Sfeas(\vec{\gamma}) = \{ \rho \in H_+ | \text{Tr}(\Gamma_k \rho)= \gamma_k \}$ and $H_+$ denotes positive semidefinite operators, $p_\text{pass}$ denotes the probability of the signal to pass sifting, and $\delta_{\text{leak}}$ is the number of bits used during error correction, per bit of raw key.

\subsection{Cascade}
In this subsection we briefly describe the error correction protocol Cascade \cite{brassard1994Secretkey}. Cascade is a simple and efficient error-correction protocol, and its principal limitation is the requirement for highly interactive communications, as compared to approaches such as LDPC codes (which suffer from a high computational cost in iterative decoding) \cite{xu2020Secure}. To understand Cascade we first look at a subprotocol called BINARY, which corrects a single error in bit strings that contain an odd number of errors. 
\begin{itemize}
	\item \textbf{BINARY: } If bit strings X and Y have odd number of errors, then Alice divides her string into halves and sends the parity of the first half to Bob. Bob divides his string the same way, and announces whether the parity of the first half is wrong, or the parity of the second half is wrong. Alice and Bob repeat the operation on the half whose parity was wrong. 
	\item The process terminates when Alice reveals the single bit which contains an error, and Bob corrects that error.  
	\item This process involves sending $\approx \log(k)$ bits from Alice to Bob, and $\approx \log(k)$ bits from Bob to Alice where $k$ is the length of the strings $X$ and $Y$. The process corrects one error. 
\end{itemize}
 \textbf{Cascade:} The Cascade protocol consists of several passes and proceeds as follows.
	\begin{enumerate}
		\item Alice and Bob divide their bit strings $X_1...X_N$ and $Y_1...Y_N$, where $N$ is the total number of sifted bits, into blocks the size of $k_1$. In pass 1, Alice and Bob reveal the parity of each block to determine the blocks with an odd number of errors. For each block with odd number of errors, Alice and Bob run BINARY to correct one error. At the end of pass 1, all blocks have even number of errors. 
		\item In any pass $i \geq 1$, Alice and Bob choose a blocksize $k_i$ and random function $f_i : [1...N] \rightarrow [1 ... N/k_i]$, which assigns each bit to a block in round $i$. The bits whose position belongs to $K^i_j = \{ l | f_i (l)=j\}$ form the $j$th block in the $i$th round. 
		\item Alice sends the parity of each block $P_{(A,i,j)} = \oplus_{l \in K^i_j} X_l $ to Bob, who computes his parity for the same block and announces it. For each block where $P_{(A,i,j)} \neq P_{(B,i,j)}$, Alice and Bob perform the following operations. 
		\begin{enumerate}
			\item Alice and Bob perform BINARY on the block defined by $K^i_j$ and correct one error, say at position $l$.  Now, all blocks in previous rounds which contained $l$ have an odd number of errors. In this way, a single error corrected in each block in later rounds leads to the identification of several error-containing blocks in earlier rounds. Let the set of such blocks be $\mathcal{K}$.  
			\item Alice and Bob choose the smallest block from $\mathcal{K}$ and run BINARY to correct one error. They again compute the set of blocks containining an odd number of errors $\mathcal{K}$. This process is repeated until $\mathcal{K}$ contains no blocks.
		\end{enumerate}
	\item At the end of pass $i$, all blocks generated in all rounds contain an even number of errors. Alice and Bob then move to the next pass.
	\end{enumerate}
\begin{remark}  The main ingredient of Cascade that we will use is the fact that for every parity bit Alice sends to Bob, Bob sends the corresponding parity bit to Alice.  There are several variants of the Cascade protocol, which vary in the manner in which blocks are created, blocksizes used, and number of passes. Such variations do not change the fact that Bob announces the same set of parities as Alice, and thus our claims will hold for all such variants. \label{remark_cascade}  \end{remark}
We note that the details of the blocks generated in a given pass have to be communicated between Alice and Bob. However, the blocks are generated randomly and independent of the QKD protocol. Therefore, the act of communicating the details of these blocks does not provide any additional information to Eve about the key \cite{Scarani2009}.

\section{Using Cascade in QKD protocols}
\label{sec:usingcascade}
\subsection{The Problem}
In the original proposal for Cascade \cite{brassard1994Secretkey}, an analytical upper bound $\delta^\text{A}_\text{leak}$ on the number of bits sent from Alice to Bob per bit of raw key is obtained. In an actual experiment, an upper bound $\delta^\text{A}_{\text{leak}}$ can also be chosen empirically by running multiple iterations of Cascade for the expected error rate. 
 For the purposes of this work, it does not matter how $\delta^\text{A}_{\text{leak}}$ is obtained. For convenience, we will denote the upper bound as $\delta^\text{A}_{\text{leak}} =  f h(e)$, where $e$ is the error-rate in the raw key, $h$ is the binary entropy function, and $f$ is a number that denotes the efficiency. Typical values of $f$ for Cascade are between $1$ and $1.5$, and can be found in \cite{brassard1994Secretkey,mao2021High,elkouss2009Efficient,martinez-mateo2015Demystifying}.
  
The original Cascade paper \cite{brassard1994Secretkey} only provides an upper bound on the number of bits sent from Alice to Bob, i.e $\delta^\text{A}_\text{leak}$, and defines `efficiency' of Cascade as the ratio of the actual number of bits per signal sent from Alice to Bob, and $h(e)$, where $e$ is the error rate and $h$ is the binary entropy function.

Therefore, it has been erroneously assumed that $\delta^\text{A}_\text{leak}$ is the true value of $\delta_{\text{leak}}$ in Eq.~\eqref{eq:basickey rate} when Cascade is used in QKD. It is assumed incorrectly that 
\begin{equation}
		\label{eq:incorrect_cascade}
		R_{\text{incorrect}} = \min_{\rho \in \Sfeas(\vec{\gamma})}	 S(Z| E \tilde{A} \tilde{B} ) - p_{\text{pass}} \times \delta^\text{A}_\text{leak}
\end{equation}
is the expression for the key rate. However, \textit{all classical communication} must be assumed to be known to Eve, and the above equation does not account for the communication from Bob to Alice during Cascade. In fact, since Bob's data is correlated with that of Alice, it is entirely possible for Bob's communication to leak additional information about Alice's raw key to Eve.

\subsection{A Naive Approach}

One naive approach to fix Eq.~\eqref{eq:incorrect_cascade} is to include $\delta^\text{B}_{\text{leak}}$, an upper bound on the number of bits leaked during the Bob to Alice communication, in $\delta_{\text{leak}}$ in Eq.~\eqref{eq:basickey rate}. Therefore, a naive, but correct expression for key rate would be
\begin{equation} 
	\label{eq:naive_cascade}
	R_{\text{naive}} = \min_{\rho \in \Sfeas(\vec{\gamma})}	 S(Z| E \tilde{A} \tilde{B} ) - p_{\text{pass}} \times (\delta^\text{A}_\text{leak} + \delta^\text{B}_\text{leak}).
\end{equation}
Here $\delta^\text{A}_\text{leak}$ can be replaced with $f h(e)$. Cascade requires Bob to send a bit to Alice for every bit Alice sends to Bob. Therefore  $\delta^\text{B}_\text{leak}=\delta^\text{A}_\text{leak}$ , leading to 
$(\delta^\text{A}_\text{leak} + \delta^\text{B}_\text{leak}) = 2 f h(e) $, which doubles the cost of error correction. Using this value in $\delta_\text{leak }$ in Eq.~\eqref{eq:naive_cascade} will yield valid key rates. However, the values obtained will be far worse than the ones obtained for any one-way error correction protocol, therefore making Cascade typically unsuitable for QKD.

\subsection{Our Solution}
\label{subsec:our_solution}
We show that one can do better than Eq.~\eqref{eq:naive_cascade}. Recalling Remark \ref{remark_cascade}, we note that the communication from Bob to Alice can be computed from two pieces of information: (1) the communication from Alice to Bob, and (2) the knowledge of the location of errors $W_i=X_i\oplus Y_i$ for each bit of Alice and Bob's data. This is because for any parity bit $P_{(A,i,j)}$ sent by Alice during Cascade, Bob sends a bit $P_{(B,i,j)}$ that is the parity of the \textit{same set of bits} of his data. Therefore $P_{(A,i,j)} \oplus P_{(B,i,j)} = \sum_{l \in K^i_j} X_l \oplus Y_l = \sum_{l \in K^i_j} W_l$, where we recall that $K^i_j$ is the set of positions of the bits that made up the $j$th block in the $i$th pass of Cascade. 

This property implies that a modified protocol that leaks all communication from Alice to Bob, and additionally leaks $W$, \textit{can only leak more (or equal) information} to Eve than Cascade. Thus, to lower bound the key rate for a QKD protocol using Cascade, we can lower bound the key rate for the QKD protocol that announces $W$, and only involves Alice to Bob part of the communication from Cascade. Therefore, we can compute 
\begin{equation} 
	\label{eq:improved_cascade}
	R = \min_{\rho \in \Sfeas(\vec{\gamma})}	 S(Z| E  W \tilde{A} \tilde{B} ) - p_{\text{pass}} \times \delta^\text{A}_\text{leak} 
\end{equation}
as a valid key rate for any QKD protocol using Cascade.
 
 We note that Eq.~\eqref{eq:incorrect_cascade} will always produce a key rate that is greater than or equal to the one produced from Eq.~\eqref{eq:improved_cascade}, since $S(Z|E\tilde{A} \tilde{B} ) \geq S(Z|EW\tilde{A} \tilde{B} )$ from subadditivity. We show that both the equality and inequality can occur, therefore proving that using Eq.~\eqref{eq:incorrect_cascade} can produce key rates that are not justified. In some cases, we obtain equality, which indicates that announcing $W$ gives Eve no new information. In such cases, although the valid key rate does not change, the argument that properly address the communication from Bob to Alice is lacking in the literature and is provided by this work.

Note that a valid key rate for \textit{any} QKD protocol involving Cascade can be obtained by considering the additional announcement of the location of errors  along with the Alice to Bob communication, as explained above. However, depending on the choice of proof technique, analyzing the protocol with the additional announcement may or may not be straightforward, and should be done carefully. In this work, we demonstrate our solution in the case of proof techniques that analyze single round entropic quantities.

\subsection{Computing Key rates}
 We are interested in the difference between the incorrect key rate from Eq.~\eqref{eq:incorrect_cascade} and the key rate from our proposed approach from Eq.~\eqref{eq:improved_cascade}. Therefore we compute two quantities,  $F =\min_{\rho \in \Sfeas(\vec{\gamma})}	 S(Z| E \tilde{A} \tilde{B} )$ and $F^\prime =  \min_{\rho \in \Sfeas(\vec{\gamma})}	 S(Z| E  W \tilde{A} \tilde{B} )$. 
	
We use the numerical framework from \cite{Winick2018} to perform these optimizations. This framework equivalently describes the steps in the QKD protocol via Kraus operators $\{K_i\}$, which represent measurements, announcements and sifting done by Alice and Bob, and  $\{Z_j\}$ which implement a pinching channel on the key register. Our solution, which requires the computation of $F^\prime$ instead of $F$ can be easily implemented by a suitable change in the Kraus operators for the optimization problem for $F$. 
	
The numerical framework equivalently describes the optimization problem for $F$ as
	\begin{equation}
		F = \min_{\rho \in \Sfeas(\vec{\gamma})}	 f(\rho),
	\end{equation}
	where 
	\begin{equation}
		\begin{aligned}
			f(\rho) &= D (\mathcal{G} (\rho) || \mathcal{Z} (\mathcal{G}(\rho) ) ), \\
			\mathcal{G}(\rho) &= \sum_i K_i \rho K_i^\dagger, \\
			\mathcal{Z} (\mathcal{G}(\rho) ) &= \sum_j Z_j \mathcal{G} (\rho) Z_j^\dagger,
		\end{aligned}
	\end{equation}
	and where $D(X || Y) = \text{Tr} (X \log (X) ) - \text{Tr} (X \log (Y))$ is the quantum relative entropy with $\log$ as the matrix logarithm. 
	
 Let $\alpha$, $\beta$ be announcements (such as basis choice) made by Alice and Bob, and let Alice and Bob's POVMs be given by $P^A = \{P^A_{(\alpha,x)}\}$, and $P^B = \{P^B_{(\beta,y)}\}$, where $x,y$ denote bits that represent measurement outcomes. Let $\mathbf{A}$ be the set of announcements $(\alpha,\beta)$ that are kept after sifting. Furthermore, let $r(\alpha,\beta,x)$ be the keymap Alice implements. Then, the Kraus operators for Eq.~\eqref{eq:incorrect_cascade} are given by
\begin{equation} \label{eq:generic_kraus}
	\begin{aligned}
	K_{\alpha,\beta}& =\sum_{x,y} \ket{r(\alpha,\beta,x) }_Z \otimes \sqrt{P^A_{(\alpha,x)} \otimes P^B_{(\beta,y)} }\\
	& \otimes \ket{x}_{X} \otimes \ket{y}_{Y} \otimes \ket{\alpha,\beta}_{\tilde{A} \tilde{B}},
	\end{aligned}
\end{equation}
and the set of operators generating the $\mathcal{G}$ map is given by $\{K_i\} = \{ K_{\alpha,\beta} | (\alpha,\beta) \in \mathbf{A} \}$  \cite{Winick2018}. 
 The $\mathcal{Z}$ map is implemented by operators $\{ Z_i\} $ given by $Z_i = \ket{i} \bra{i}_Z \otimes I_{ABX Y \tilde{A} \tilde{B}}$. Notice that the output state $\mathcal{G}(\rho)$ is classical in $\alpha,\beta$, which reflects the fact that the basis choices are announced and known to Eve. 

 To compute $F^\prime$, we must include an additional announcement that announces $w=x \oplus y$. This is implemented by 
\begin{equation} \label{eq:newkraus}
	\begin{aligned}
	K^\prime_{\alpha,\beta, w} &=\sum_w \sum_{\substack{x,y \\ x\oplus y = w}} \ket{r(\alpha,\beta,x) }_Z \otimes \sqrt{P^A_{(\alpha,x)} \otimes P^B_{(\beta,y)} }\\
	& \otimes \ket{x}_X \otimes \ket{y}_Y  \otimes \ket{\alpha,\beta}_{\tilde{A} \tilde{B} } \otimes \ket{w}_{W},	
	\end{aligned}
\end{equation}
where the set of operators generating the new $\mathcal{G}^\prime$ map can be given by $\{K^\prime_i\} = \{ K^\prime_{\alpha,\beta, w} | (\alpha,\beta) \in \mathbf{A} \}$.  The $\mathcal{Z}^\prime$ map is implemented by $\{ Z^\prime_i\} $ given by $Z^\prime_i = \ket{i} \bra{i}_Z \otimes I_{ABX Y \tilde{A} \tilde{B}W}$.

 In the remainder of this paper, we compute both $F= \min_{\rho \in \Sfeas(\vec{\gamma})}	 f(\rho)$ and $F^\prime = \min_{\rho \in \Sfeas(\vec{\gamma})}	 f^\prime(\rho)$ for the various implementations of the BB84 protocol. If we find that $F=F^\prime$, then this indicates that the previous analysis of Cascade is wrong but gives correct answers. In this case, Eqs.~\eqref{eq:incorrect_cascade} and \eqref{eq:improved_cascade} will give identical key rates.
If we find that $F > F^\prime$, then this indicates that the previous analysis was wrong and gave incorrect answers. The difference between the key rates obtained from Eqs.~\eqref{eq:incorrect_cascade} and ~\eqref{eq:improved_cascade} is equal to $F-F^\prime$.  Note that $F-F^\prime$ is the difference in the key rates measured in terms of secure key bits per signal sent. The difference in key rates in terms of secure key bits per second is given by $R_r(F-F^\prime)$, where $R_r$ is the repetition rate.

 Note that the above formulation applies to situations where Alice and Bob generate a bit string from their measurements ($x,y,x \oplus y$ are bits). Events such as no-detection either need to be discarded during sifting, or should be mapped to bits. This assumption is necessary to use Cascade, since it is a protocol that corrects errors in \textit{bit strings}. We also remark that since many finite-size key rate analyses involve the optimization of the same objective function ($F$) over different constraints, our solution can be easily applied to such finite-size analyses as well, by simply changing $F$ to $F^\prime$.

\section{Qubit BB84}
\label{sec:qubit_bb84}

\subsection{Protocol Description}
In this section, we present our results for the standard qubit-based BB84 protocol, where Alice prepares each of the four signal states $\{ \ket{0}, \ket{1}, \ket{+}, \ket{-} \}$ with equal probability, and Bob chooses the $Z$ or $X$ basis with equal probability.
Alice and Bob then implement an acceptance test on their observed statistics.
If the protocol accepts, Alice and Bob announce their basis, and throw away signals where they measured in different basis. Alice maps her measurement outcome to the raw key, and then proceeds to perform error-correction (Cascade) and privacy amplification. For the descriptions of the exact Kraus operators of the protocol, we refer the reader to Appendix \ref{appendix:qubitBB84}.

\begin{table}
	\centering
	
	\begin{tabular}{|c|ccccc|}
		
		\hline
		\multicolumn{6}{|c|}{Bob Measures} \\
		\hline
		&  & H & V & + & - \\		\hline
		& H & \mk{$ \gamma_{HH}$ }  &  \mk{$\gamma_{HV}$} & $\gamma_{H+}$ & $\gamma_{H-}$  \\
		
		Alice&  V & \mk{$\gamma_{VH}$} & \mk{$\gamma_{VV}$} & $\gamma_{V+}$  & $\gamma_{V-}$  \\
		
		Sends&+& $\gamma_{+H}$ & $\gamma_{+V}$ & \mk{$\gamma_{++}$ } & \mk{$\gamma_{+-}$}  \\
		
		& -& $\gamma_{-H}$ & $\gamma_{-V}$ & \mk{$\gamma_{-+}$} & \mk{$\gamma_{--}$ }\\
		\hline
	\end{tabular}
	\caption{Format of fine-grained statistics obtained for qubit BB84. The rows denote the state sent by Alice, and the columns denote the measurement outcome measured by Bob. $H$ and $+$ correspond to measurement outcome $0$, while $V$ and $-$ correspond to measurement outcome $1$. $\gamma_{x,y}$ denotes the probability of Alice obtaining outcome $x$ and Bob obtaining outcome $y$.}
	\label{table:qubit_constraints}
\end{table}

Alice and Bob obtain statistics shown in Table \ref{table:qubit_constraints} during the acceptance test. There are a variety of ways they can use these statistics to perform the acceptance test. We use the phrase ``fine-grained constraints'' to refer to the case where all the entries in Table \ref{table:qubit_constraints} are used for the acceptance test, and therefore in the constraints for $\Sfeas(\vec{\gamma})$. We use ``sifted fine-grained'' to refer to the case where only the entries marked in red are used. We use ``coarse-grained" constraints to refer to the case where only the (unnormalized) QBER and Gain constraints for each basis (given by $\text{Q}_Z = \gamma_{HV}+ \gamma_{VH}$, $\text{Q}_X = \gamma_{+-} + \gamma_{-+}$, $\text{gain}_Z = \gamma_{HH}+ \gamma_{HV}+ \gamma_{VH}+ \gamma_{VV}$, and $\text{gain}_X = \gamma_{++}+ \gamma_{+-}+ \gamma_{-+}+ \gamma_{--}$)  are used. The gain sets the constraints on the probability of choosing each basis for measurement, while the QBER in each basis sets the constraints on the observed error-rate. We note that this is a departure from the nomenclature of \cite{wang2021Numerical}, where the ``coarse-grained'' case refers to the ``sifted fine-grained case'' as defined above.  Additionally, we use the constraints from source-replacement that characterize Alice's system for prepare and measure protocols.

We consider a channel with misalignment and depolarization to compute statistics in Table \ref{table:qubit_constraints}, which is described in Appendix \ref{appendix:qubitBB84}. We also consider a channel model that includes a  ``replacement'' channel $\Phi_{\text{replace}}$, that replaces the state leaving Alice's lab with the fixed signal state $\ket{0}$ ($\ket{H}$) with probability $\lambda=0.2$. The output of the replacement channel is then sent through a channel with misalignment and depolarization. The replacement channel is interesting because it breaks symmetries in the observed statistics. 
If the replacement channel is also included in the channel model, then the new statistics can be obtained by replacing each row $\vec{\gamma}_i$ of Table \ref{table:qubit_constraints} by $(1-\lambda)\vec{\gamma}_i + \lambda\vec{\gamma}_H$ (since Alice sends each state with equal probability).

\subsection{Reduction to Bell-diagonal states}
\label{subsec:reduction}
In certain cases, the optimization of $f(\rho)$ over all states $\rho$ in $\Sfeas(\vec{\gamma)}$, can be reduced to that over all Bell-diagonal states in $\Sfeas(\vec{\gamma})$, denoted by $\Sfeas_\text{bell} (\vec{\gamma})$. That is, it can be shown that
\begin{equation} \label{eq:reduction}
	\min_{\rho \in \Sfeas_\text{bell}(\vec{\gamma}) } f(\rho) = \min_{\rho \in \Sfeas (\vec{\gamma})} f(\rho).
\end{equation} 
For Bell-diagonal states shared between Alice and Bob, Eve's state $E$ is always block-diagonal in the parities $W$ (see Appendix \ref{appendix:bell}), and therefore the additional announcement $W$ gives Eve no new information. In such cases, $F=F^\prime$.

There are several such arguments in the literature, which are identical at their core, but differ only in details of the protocols (such as type of constraints, number of basis used for key generation, and type of classical processing). Ref.~\cite{ferenczi2011Symmetries} proves Eq.~\eqref{eq:reduction} for the case where $f(\rho)$ is the total key rate including the information leakage term, and the states are constrained in $\Sfeas$ only by the average QBER over all bases. The analysis is done for $d$ dimensional systems in general. Ref.~\cite{Watanabe2007} proves Eq.~\eqref{eq:reduction} for BB84 and six-state protocols, where $f(\rho)$ is the uncertainity of Eve about the raw key (with a modified classical processing), and the states are constrained in $\Sfeas$ by each individual QBER, but only the Z basis is used for key generation. Ref.~\cite{tupkary2022Improved} generalizes this to a wider variety of classical processing in $f(\rho)$, while still constraining $\Sfeas$ with separate QBERs, but using only the Z basis for key generation. In this work, we will attempt to present a coherent unified picture of all such arguments for the convenience of the reader. We also point out how symmetry in \textit{observed values} can help in proving the reduction to Bell-diagonal states.

We start by defining the ``twirling map" \cite{bennett1996Mixed} as
 \begin{equation}
		\mathcal{T}(\rho)= \frac{1}{4} \sum_{i=1}^4 \rho_i = \frac{1}{4} \sum_{i=1}^4 (\sigma_i \otimes \sigma_i) \rho (\sigma_i \otimes \sigma_i)^\dagger.
	\end{equation}
	where $\sigma_i$ for $i \in \{0,1,2,3\}$ denotes the identity and the Pauli X,Y and Z operators. $\mathcal{T}(\rho)$ is often referred to as the ``twirled'' state, and can be shown to be always Bell-diagonal \cite{ferenczi2011Symmetries,bennett1996Mixed}. The proof of Eq.~\eqref{eq:reduction} now proceeds in two steps.

\noindent
\textit{Step 1 :} It is shown that 
	\begin{equation} \label{eq:first_condition}
		f(\mathcal{T}(\rho)) \leq f(\rho) \quad \forall \rho.
	\end{equation} 
\textit{Step 2 :} It is shown that 
	\begin{equation} \label{eq:second_condition}
		\rho \in \Sfeas \implies \mathcal{T}(\rho) \in \Sfeas_{\text{bell}}.
	\end{equation}
The proof of Eq.~\eqref{eq:reduction} is then as follows : Clearly $\min_{\rho \in \Sfeas_\text{bell} (\vec{\gamma}) } f(\rho) \geq \min_{\rho \in \Sfeas(\vec{\gamma})} f(\rho)$, since $\Sfeas_\text{bell} (\vec{\gamma}) \subseteq \Sfeas (\vec{\gamma}) $. To show the other direction of the inequality, let $\rho^*$ be the state that achieves the minimization on the right hand side of Eq.~\eqref{eq:reduction}. Then, from Eq.~\eqref{eq:first_condition}, we obtain $f(\mathcal{T}(\rho^*)) \leq f(\rho^*)$. From Eq.~\eqref{eq:second_condition}, we know that $\mathcal{T}(\rho^*) \in \Sfeas_\text{bell} (\vec{\gamma})$. Therefore, $	\min_{\rho \in \Sfeas_\text{bell} (\vec{\gamma}) } f(\rho) \leq f(\mathcal{T}(\rho^*)) \leq \min_{\rho \in \Sfeas (\vec{\gamma})} f(\rho)$.

Thus, to obtain Eq.~\eqref{eq:reduction} for a protocol of interest, one must show the validity of Eqs.~\eqref{eq:first_condition} and \eqref{eq:second_condition}. We prove that Eq.~\eqref{eq:first_condition} holds for qubit protocols where key generation is done in all the X, Z (and if applicable Y) basis in Appendix \ref{appendix:twirling}. Thus, to reduce the optimization to Bell-diagonal states and obtain $F=F^\prime$, we now only need to check the validity of Eq.~\eqref{eq:second_condition}. This has to be considered separately for every choice of constraints, and observed values, and is done in the next section.

\subsection{Results}

We numerically check the difference between $F$ and $F^\prime$. The results are summarized in Table \ref{table:qubitbb84_results}. Since the numerical method is capable of producing both an upper bound and lower bound for $F$ and $F^\prime$, it is straightforward to determine when $F>F^\prime$. We claim $F=F^\prime$ when the bounds for $F$ and $F^\prime$ overlap. In some cases, $F=F^\prime$ can be analytically argued, by proving the validity of Eq.~\eqref{eq:second_condition} (see Sec. \ref{subsec:reduction}), as we do below. 

To check the validity of Eq.~\eqref{eq:second_condition}, we need to look at the constraints that define $\Sfeas(\vec{\gamma})$. There are two types of constraints. First, we have the constraint obtained from the source-replacement scheme, which is of the form $\text{Tr}_B(\rho) = \sigma_A$ and represents the fact that in prepare and measure protocols, Alice's state is known and never leaves her lab. For the qubit BB84 protocol, one can take $\sigma_A = I_A /2$ (see Appendix. \ref{appendix:qubitBB84}). Since $\text{Tr}_B(\rho_{AB}) = I_A / 2$ is always true for a Bell-diagonal state, this constraint is always satisfied by $\mathcal{T}(\rho)$.

The remaining constraints are obtained from the acceptance test, and are of the form $\text{Tr}(\Gamma_k \rho ) = \gamma_k$. Thus, to check the validity of Eq.~\eqref{eq:second_condition}, we check whether $\text{Tr}(\Gamma_k \mathcal{T}(\rho)) = \text{Tr}(\mathcal{T}^\dagger(\Gamma_k) \rho ) =\gamma_k \quad \forall \rho$.
\begin{table*}[t!]
	
	\begin{tabular}{|c|c|c|c|c|c|}
		\hline
		\multicolumn{6}{|c|}{Channel} \\
		\hline
		 & & Misalignment  & Depolarization  & Misalignment + & Misalignment+ \\ 
		 & & & & Depolarization & Depolarization+ \\
		 & & & & & $\Phi_\text{replace}$ \\
		\hline
		& Coarse-grained & $=$ & $=$ & $= $& $=$ \\
		Constraints& Sifted fine-grained & $=$ & $=$ & $=$& $>$ \\
			& Fine-grained & $=$ & $=$ & $>$ & $>$ \\
		\hline
	\end{tabular}
	\caption{Relation between $F$ and $F^\prime$ for Qubit BB84 protocol. Results are based upon the upper and lower bound to the optimization obtained from the numerical method \cite{Winick2018}. All equality cases can be explained by arguing a reduction to Bell-diagonal states.}
	\label{table:qubitbb84_results}
	\end{table*}

\begin{figure}
	\includegraphics[width=\linewidth]{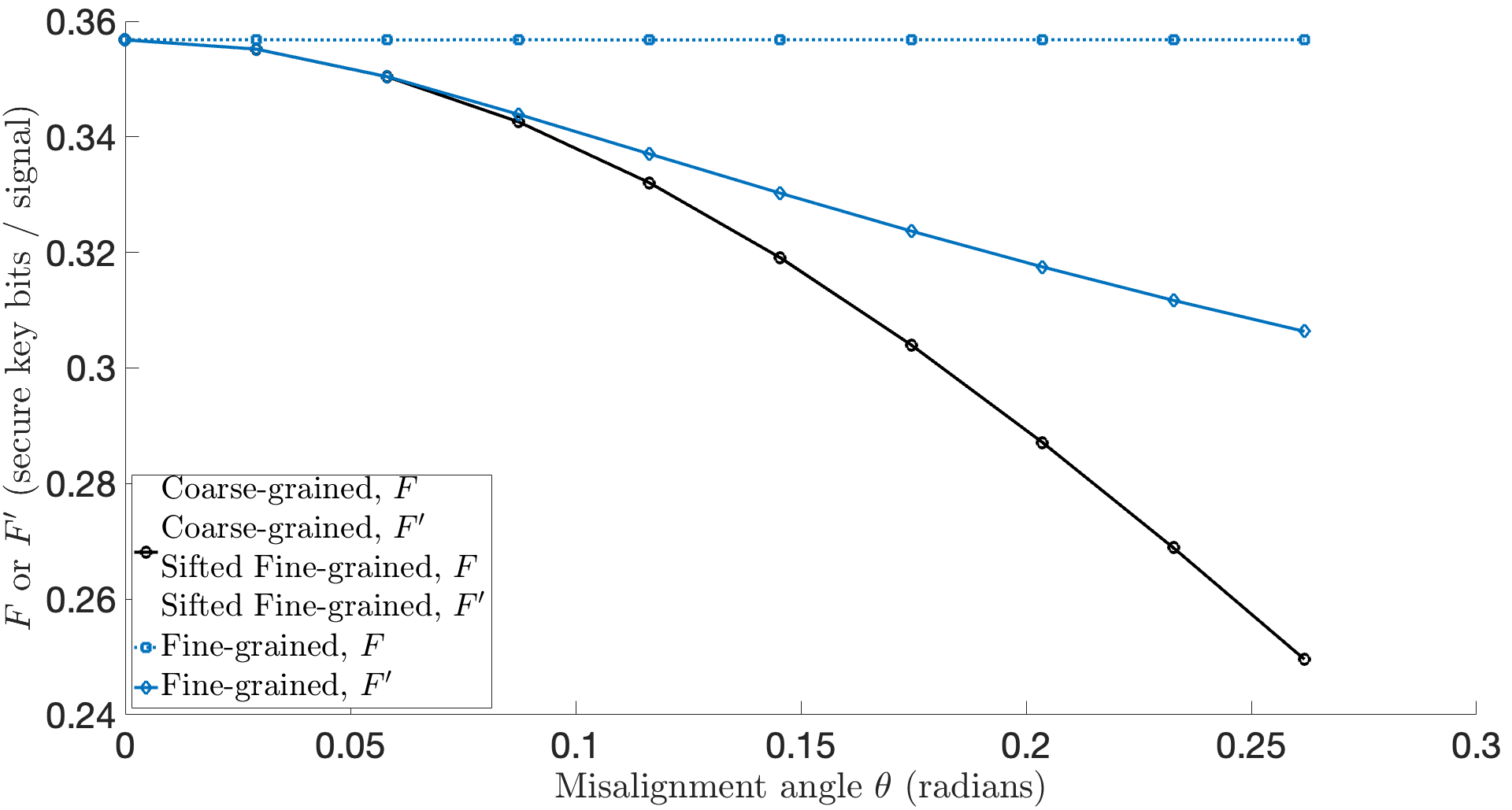}
	\caption{$F,F^\prime$ for a channel with misalignment and depolarization. We find that $F=F^\prime$ for coarse-grained and sifted fine-grained constraints, while $F>F^\prime$ for fine-grained constraints. The plot corresponds to depolarization probability $q=0.1$, and is plotted against the misalignment angle $\theta$.}
	\label{fig:misalign_depol}
\end{figure}

\begin{figure}
	\includegraphics[width=\linewidth]{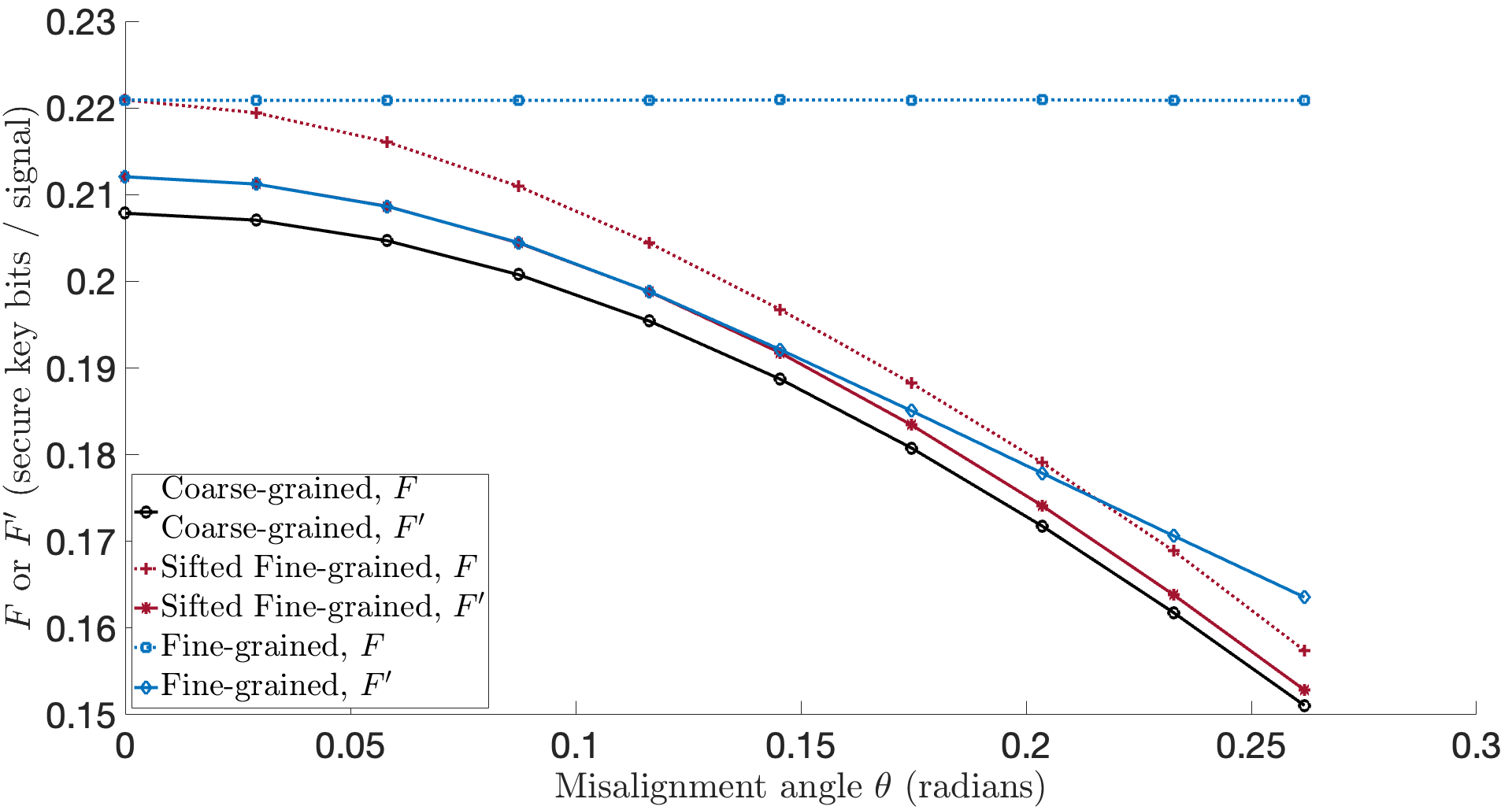}
	
	\caption{$F,F^\prime$ for a channel with misalignment, depolarization and replacement channel. We find that $F=F^\prime$ for coarse-grained, while $F>F^\prime$ for sifted fine-grained and fine-grained constraints. The state leaving Alice's lab is replaced with the signal state corresponding to $H$, with probability $\lambda = 0.2$. The plot corresponds to depolarization probability $q=0.1$, and is plotting against the misalignment angle $\theta$. }
	\label{fig:misalign_depol_perturb}
\end{figure}

\begin{itemize}
	\item \textbf{Coarse-grained statistics :} In this case, $\Gamma_k$ is the POVM element corresponding to QBER and gain in each basis. It is therefore easy to check with a simple calculation that $\mathcal{T}^\dagger (\Gamma_k) = \Gamma_k$ which implies $\text{Tr}(\Gamma_k \mathcal{T}(\rho)) = \text{Tr}(\Gamma_k \rho)$. Thus, for coarse-grained constraints the state shared between Alice and Bob can be assumed to be Bell-diagonal, and $F=F^\prime$. This explains the coarse-grained row in Table \ref{table:qubitbb84_results}.

	\item \textbf{Sifted Fine-grained statistics:} Let us turn to the case of ``sifted fine-grained'' constraints. For the $Z$ basis, let the POVMs that make up the constraints be given by $\Gamma_{HH},\Gamma_{HV}, \Gamma_{VH}, \Gamma_{VV}$ (with similar expressions for the $X$ basis). In general, these POVMs are not invariant under $\mathcal{T}^\dagger$, and thus $\text{Tr}(\Gamma_k \mathcal{T}(\rho)) \neq \text{Tr}(\Gamma_k \rho)$. However, one can find that 
\begin{equation} \label{eq:symmetry_POVMS}
	\begin{aligned}
	\mathcal{T}^\dagger ( \Gamma_{HH}) = \mathcal{T}^\dagger( \Gamma_{VV}  ) &= \frac{1}{2} (\Gamma_{HH}+\Gamma_{VV}), \\
	\mathcal{T}^\dagger(  \Gamma_{HV}) = \mathcal{T}^\dagger ( \Gamma_{VH}  ) &= \frac{1}{2} ( \Gamma_{VH}+ \Gamma_{HV}), \\
	\mathcal{T}^\dagger (  \Gamma_{++} ) = \mathcal{T}^\dagger( \Gamma_{--} ) &= \frac{1}{2}( \Gamma_{++} + \Gamma_{--} ), \\
	\mathcal{T}^\dagger ( \Gamma_{+-} ) = \mathcal{T}^\dagger( \Gamma_{-+} )  &= \frac{1}{2}( \Gamma_{+-}+ \Gamma_{-+} ).
	\end{aligned}
\end{equation}
	
	Therefore, in this case, one can claim a reduction to Bell-diagonal as long as the statistics obey certain symmetries. That is, if one obtains statistics satisfying $\gamma_{HH}=\gamma_{VV}, \gamma_{HV}=\gamma_{VH}, \gamma_{++}=\gamma_{--}, \gamma_{+-}=\gamma_{-+}$, then
	\begin{equation}
		\begin{aligned}
			\text{Tr}(\Gamma_k \rho) = \gamma_k \implies &\text{Tr}(\Gamma_k \mathcal{T}(\rho) ) = \gamma_k, \\
			 \text{even when } & \mathcal{T}^\dagger (\Gamma_k) \neq \Gamma_k.
			\end{aligned}
	\end{equation}
	
	The statistics obey this symmetry when the channel consists of any combination of loss and misalignment, and therefore for these channel models we again obtain $F=F^\prime$ due to the reduction to Bell-diagonal states. Introducing the additional replacement channel $\Phi_\text{replace}$ destroys this symmetry, and we obtain $F\neq F^\prime$. This explains the sifted fine-grained row in Table \ref{table:decoybb84_results}.

	 \item \textbf{Fine-grained statistics :} In this case, in addition to Eq.~\eqref{eq:symmetry_POVMS}, it is possible to show that each POVM in the off-diagonal block of Table \ref{table:qubit_constraints} is mapped to the same off-diagonal block by $\mathcal{T}^\dagger$. That is, 
	 \begin{equation}
	 	\begin{aligned}
	 		\mathcal{T}^\dagger( \Gamma_{H+})  &=	\mathcal{T}^\dagger(\Gamma_{H-}) = 	\mathcal{T}^\dagger(\Gamma_{V+}) = 	\mathcal{T}^\dagger(\Gamma_{V-}) \\
	 		&= \frac{1}{4} (\Gamma_{H+}+\Gamma_{H-}+\Gamma_{V+}+\Gamma_{V-}) \\
	 		\mathcal{T}^\dagger( \Gamma_{+H}) &= 	\mathcal{T}^\dagger(\Gamma_{+V}) = 	\mathcal{T}^\dagger(\Gamma_{-H})  =	\mathcal{T}^\dagger(\Gamma_{-V}) \\
	 		 &= \frac{1}{4} (\Gamma_{+H}+\Gamma_{+V}+\Gamma_{-H}+\Gamma_{-V})	 		 
	 		\end{aligned}
	 \end{equation}
	 
	 That is, if one obtains statistics satisfying $\gamma_{HH}=\gamma_{VV}, \gamma_{HV}=\gamma_{VH}, \gamma_{++}=\gamma_{--}, \gamma_{+-}=\gamma_{-+}$ along with $\gamma_{H+} = \gamma_{H-} = \gamma_{V+} = \gamma_{V-}$, and $\gamma_{+H} = \gamma_{+V} = \gamma_{-H} = \gamma_{-V}$, then we again can claim that
	 \begin{equation}
	 	\begin{aligned}
	 		\text{Tr}(\Gamma_k \rho) = \gamma_k \implies &\text{Tr}(\Gamma_k \mathcal{T}(\rho) ) = \gamma_k, \\
	 		 \text{even when } &
	 		\mathcal{T}^\dagger (\Gamma_k) \neq \Gamma_k.
	 	\end{aligned}
	 \end{equation}
	  This is the case when the channel only contains depolarization.
	
	 For only misalignment, it has already been shown that fine-grained constraints allow us to show that Eve factors off and holds a state that is independent of the Alice-Bob state \cite{wang2021Numerical}. Since Eve's quantum system factors off, the $F=F^\prime$ follows from the fact that $W$ and $Z$ are independent random variables for each basis, i.e $ S(Z | \tilde{A} \tilde{B} W) = S(Z | \tilde{A} \tilde{B} )$. For the remaining two cases, we find that the optimal values of the two objective functions are unequal, and no reduction to the Bell-diagonal case is possible.   This explains the fine-grained row of Table \ref{table:qubitbb84_results}.
\end{itemize}
We plot $F, F^\prime$ corresponding to the last two columns of Table \ref{table:decoybb84_results} in Figs. \ref{fig:misalign_depol}, \ref{fig:misalign_depol_perturb}. Note that we always find that the fine-grained $F(F^\prime)$ is higher than the sifted fine-grained $F(F^\prime)$, followed by the coarse-grained $F(F^\prime)$, as has already been pointed out in Ref.~\cite{wang2021Numerical}.

\section{WCP Decoy State BB84}
\label{sec:WCP_Decoy}
In this section, we present results for the WCP decoy-state BB84 protocol \cite{lim2014Concise,lo2005Decoy,rusca2018Finitekey,rice2009Numerical,hwang2003Quantum,ma2005Practical,Wang_2005} in the same manner. The key rate calculations, protocol description, channel simulation, and decoy analysis is exactly identical to the one from \cite{wang2021Numerical}. Therefore, these aspects will be only briefly described in this work. The only difference lies in the modification of Kraus operators according to Eqs.~\eqref{eq:newkraus}, and the inclusion of the replacement channel $\Phi_{\text{replace}}$ in the channel simulation.  

\subsection{Protocol Specification}
Alice prepares and sends a phase-randomized weak coherent pulse (WCP) pulse in the polarization mode  $H,V,A,D$ with equal probability, choosing to use the ``signal intensity'' with probability close to one, and some ``decoy intensities''. Bob implements passive basis choice with equal probability.  We use a squashing model on Bob's side \cite{Gittsovich2014} to describe his measurements,  and Bob's squashed POVMs can be found in Appendix \ref{appendix:decoyprotocol}.  Alice and Bob announce a small fraction of their data, and perform the acceptance test. If the protocol accepts, Alice and Bob announce basis, and throw away the signals where they measured in different basis, or where Bob got a no detection event. Alice then maps her measurement outcomes to the raw key, followed by error correction (Cascade) and privacy amplification. For the descriptions of the exact Kraus operators of the protocol, we refer the reader to Appendix \ref{appendix:decoyprotocol}.

\begin{table}[h!]
	\centering
	
	\begin{tabular}{|c|c| ccccc|}
		
		\hline
		\multicolumn{7}{|c|}{Bob Measures} \\
		\hline
		&  & H & V & + & - & $\emptyset$ \\		\hline
		& H & \mk{$ \gamma^{\mu_i}_{HH}$ }  &  \mk{$\gamma^{\mu_i}_{HV}$} & $\gamma^{\mu_i}_{H+}$ & $\gamma^{\mu_i}_{H-}$ & $\gamma^{\mu_i}_{H \emptyset}$  \\
		
		Alice&  V & \mk{$\gamma^{\mu_i}_{VH}$} & \mk{$\gamma^{\mu_i}_{VV}$} & $\gamma^{\mu_i}_{V+}$  & $\gamma^{\mu_i}_{V-}$ & $\gamma^{\mu_i}_{V \emptyset}$   \\
		
		Sends&+& $\gamma^{\mu_i}_{+H}$ & $\gamma^{\mu_i}_{+V}$ & \mk{$\gamma^{\mu_i}_{++}$ } & \mk{$\gamma^{\mu_i}_{+-}$} & $\gamma^{\mu_i}_{+ \emptyset}$   \\
		
		& -& $\gamma^{\mu_i}_{-H}$ & $\gamma^{\mu_i}_{-V}$ & \mk{$\gamma^{\mu_i}_{-+}$} & \mk{$\gamma^{\mu_i}_{--}$ } & $\gamma^{\mu_i}_{- \emptyset}$  \\
		\hline
	\end{tabular}
	\caption{Format of fine-grained statistics obtained for deocy-state BB84 protocol. The rows denote the state sent by Alice, and the columns denote the measurement outcome measured by Bob. One such table is obtained for each intensity used by Alice. $H$ and $A$ correspond to measurement outcome $0$, while $V$ and $D$ correspond to measurement outcome $1$. $\gamma^{\mu_i}_{x,y}$ denotes the probability of Alice obtaining outcome $x$ and Bob obtaining outcome $y$, given intensity $\mu_i$ was used.}
	\label{table:decoybb84_constraints}
\end{table}
The fine-grained statistics obtained by Alice and Bob are given by the Table \ref{table:decoybb84_constraints}. Again, as in Sec. \ref{sec:qubit_bb84}, we use the phrase ``fine-grained constraints'' to refer to the case where all the entries in Table \ref{table:qubit_constraints} are used for the acceptance test, ``sifted fine-grained'' when only the entries marked in red are used, and ``coarse-grained" constraints when only the (unnormalized) QBER and Gain constraints for each basis are used. Additionally, we use the constraints from source-replacement that characterize Alice's system for prepare and measure protocols. 

The exact manner in which statistics from  a channel consisting of misalignment and loss are computed is identical to the procedure described in Ref.~\cite[Appendix C]{wang2021Numerical}. We will not repeat those calculations here. We include an additional ``replacement channel'' $\Phi_{\text{replace}}$ which replaces the state leaving Alice's lab with the signal state corresponding to $H$ with probability $\lambda = 0.2$. This is interesting since it breaks symmetries in observed statistics. If we include the replacement channel, then each row  $\vec{\gamma}^{\mu_i}_j$ of Table \ref{table:decoybb84_constraints}  (computed for loss and misalignment), is replaced by $(1-\lambda)\vec{\gamma}^{\mu_i}_j + \lambda\vec{\gamma}^{\mu_i}_H$ (since Alice sends each state with equal probability).

\subsection{Results}

\begin{table*}
	\centering
	
	\begin{tabular}{|c|c|c|c|c|c|}
		\hline
		\multicolumn{6}{|c|}{Channel} \\
		\hline
		& & Loss  & Misalignment  & Loss+Misalignment & Loss+Misalignment+ \\
		& & & & & $\Phi_\text{replace}$ \\
		\hline
		& Coarse-grained & $=$ & $=$ & $= $& $=$ \\
		Constraints& Sifted fine-grained & $=$ & $=$ & $=$& $>$ \\
		& Fine-grained & $=$ & $=$ & $>$ & $>$ \\
		\hline
	\end{tabular}
	\caption{Relation between $F$ and $F^\prime$ for Decoy BB84 protocol. Results are based upon the upper and lower bound to the optimization obtained from the numerical method \cite{Winick2018}. Note that this table is similar to Table \ref{table:qubitbb84_results} obtained for qubit BB84, suggesting the fact that similar arguments can be made for understanding this table in both cases.}
	\label{table:decoybb84_results}
\end{table*}

The optimization problem for decoy protocols is solved by obtaining bounds on the single photon yields as 	
\begin{equation}
	   \gamma^{1,L}_{y | x} \leq \gamma^1_{y|x} \leq  \gamma^{1,U}_{y|x}, \quad \forall x,y 
\end{equation} 
where $\gamma^1_{y|x}$ denotes the probability of Bob obtaining outcome $y$, given Alice sent signal $x$ and $1$ photon.  One can then compute lower and upper bounds on $\gamma^1_{x,y}$, by using $\gamma^1_{x,y} = \Pr(x) \gamma^1_{y|x}$, where $\Pr(x)$ denotes the probability of Alice sending signal $x$.

The optimization problem \cite{ma2005Practical,lo2005Decoy,rice2009Numerical,wang2021Numerical} is then given by (see Appendix \ref{appendix:decoy})
\begin{equation} \label{eq:decoy_opt}
	\begin{aligned}
		F &= \min_{\rho \in \Sfeas_1^\prime } f(\rho), \\
		\Sfeas_1^\prime & = \{ \rho \in H_+ | \gamma^{1,L}_k \leq  \text{Tr}(\Gamma_k \rho)    \leq \gamma^{1,U}_k, \forall k\}
	\end{aligned}
\end{equation}
where $H_+$ denotes positive semidefinite operators,  and $\Sfeas^\prime_1$ is the set of density operators compatible with observed statistics, and $k$ depends on the exact nature of coarse-graining. 

We numerically compute the difference between $F= \min_{\rho \in \Sfeas_1^\prime } f(\rho)$ and $F^\prime =  \min_{\rho \in \Sfeas_1^\prime } f^\prime(\rho)$ for all our channel models, and various types of constraints. The results are summarized in Table \ref{table:decoybb84_results}. 

\begin{figure}
		\includegraphics[width=\linewidth]{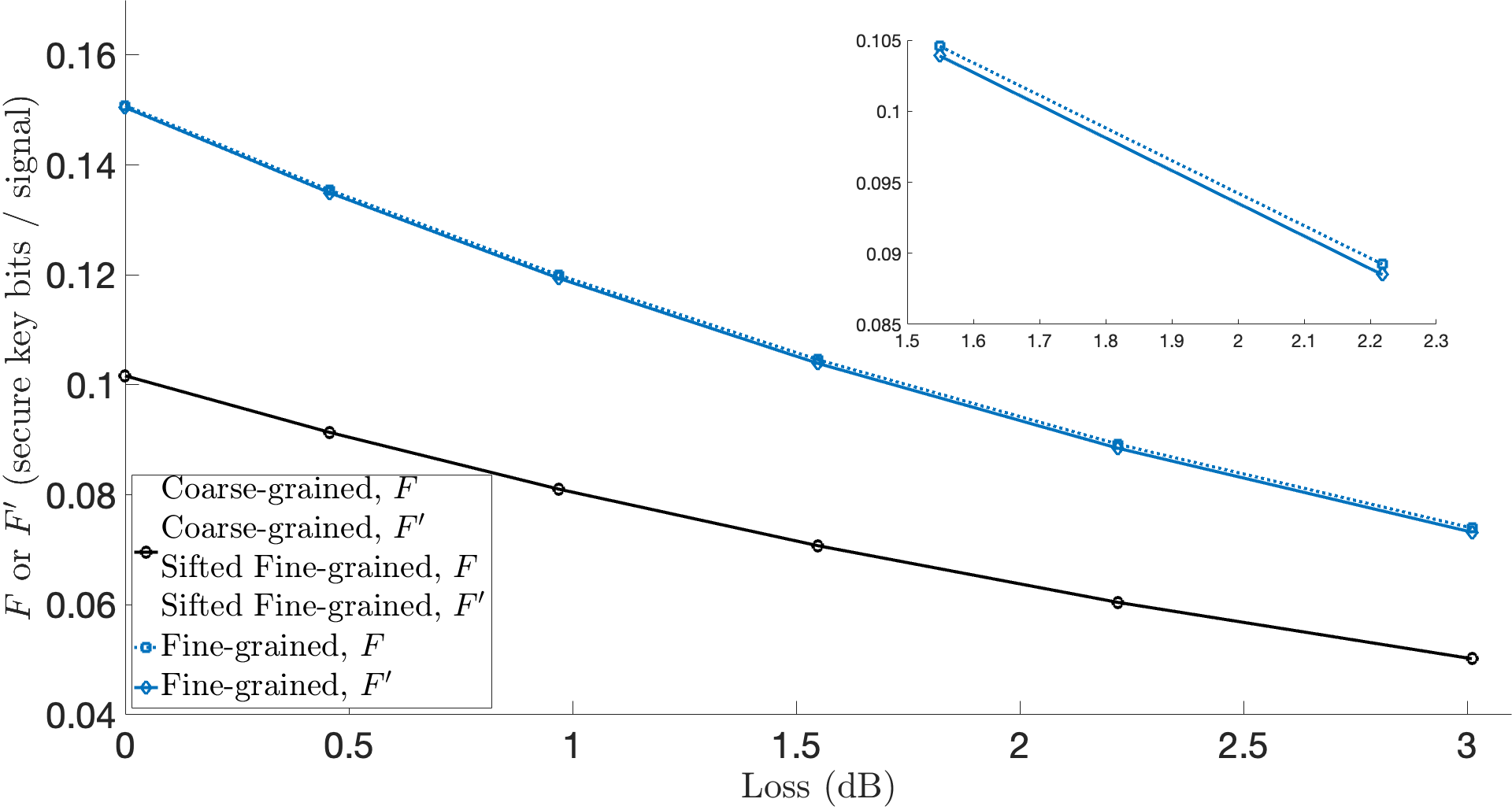}
		\caption{$F,F^\prime$ for a channel with misalignment and loss. We find that $F=F^\prime$ for coarse-grained and sifted fine-grained constraints, while $F>F^\prime$ for fine-grained constraints. The plot corresponds to a misalignment angle $\theta$ given by $\sin^2(\theta)=0.06$, and three intensities $\mu_1=0.5, \mu_2=0.1, \mu_3= 0.001$, with the first intensity used to generate the key.}
		\label{fig:misalign_loss}
	\end{figure}

\begin{figure}
		\includegraphics[width=\linewidth]{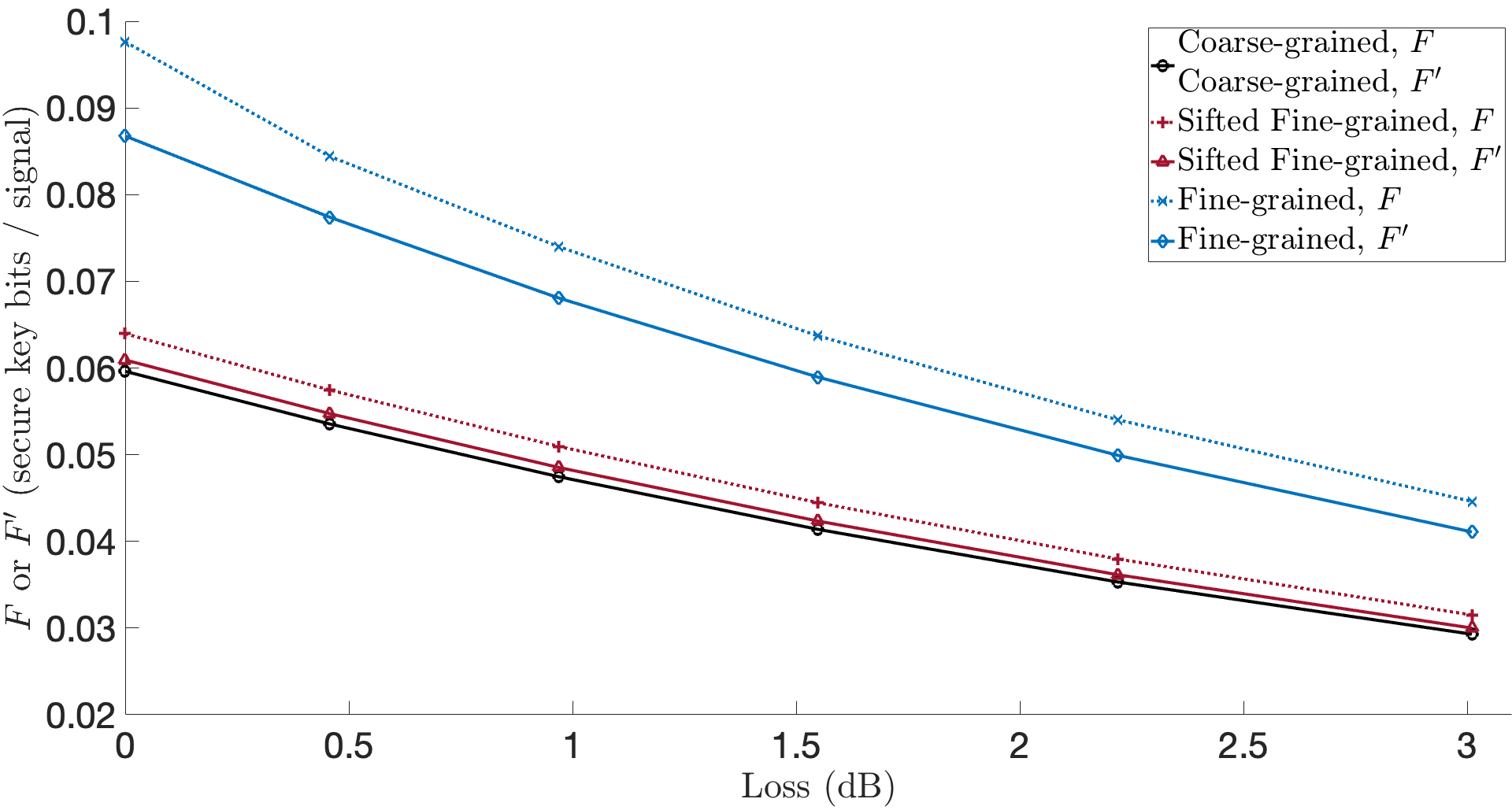}
		\caption{$F,F^\prime$ for a channel with misalignment, loss and replacement channel. We find that $F=F^\prime$ for coarse-grained, while $F>F^\prime$ for sifted fine-grained and fine-grained constraints. The state leaving Alice's lab is replaced with the signal state corresponding to $H$ with probability $\lambda=0.2$.
			The plot corresponds to a misalignment angle $\theta$ given by $\sin^2(\theta)=0.06$, and three intensities $\mu_1=0.5, \mu_2=0.1, \mu_3= 0.001$, with the first intensity used to generate the key.}
		\label{fig:misalign_loss_perturb}
	\end{figure}

Since after squashing, the single photon contribution to the objective function involves both Alice and Bob having qubits (or vacuum), our intuition from the qubit BB84 picture can be used to understand the results in Table \ref{table:decoybb84_results}. We believe a more rigorous justification can be made along the same lines as for the qubit case, however that is not a contribution of this work.  For coarse-grained constraints, we expect symmetry arguments to allow us to restrict to Bell-diagonal states, in which case announcing $W$ provides no new information to Eve. For the ``sifted fine-grained'' and ``fine-grained'' case, we expect symmetry arguments to not work in general, but to allow a restriction to Bell-diagonal states if observations are also symmetric, as seen in Sec. \ref{sec:qubit_bb84}. We plot $F, F^\prime$ corresponding to the last two columns of the Table \ref{table:decoybb84_results} in Figs. \ref{fig:misalign_loss} and \ref{fig:misalign_loss_perturb}.
	
 \textbf{Effect of zero-photon contribution: } The above analysis is done for the case where we only keep the single-photon contribution to the key in Eq.~\eqref{eq:decoy_opt} (Eq.~\eqref{eq:objfunc_diffn}). Let us consider the zero-photon contribution to the key. In this case, note that since no signal left Alice's lab, Eve cannot know anything about Alice's key bit. Therefore, the zero-photon contribution to $F$ is  equal to $p^{(0)}_{\text{pass}}$, where $p^{(0)}_{\text{pass}}$ is the probability of zero-photon event leading to detection and passing sifting. 

Moreover, if Alice sends zero photons, the state giving rise to Bob's detection must be assumed to be known to Eve. Therefore Eve has perfect knowledge of Bob's data. In this case, if $W$ is announced, Eve has perfect knowledge of Alice's data as well. Therefore, the zero-photon contribution to $F^\prime$ is  \textit{always zero}.
	 
 Therefore, in this case $F > F^\prime$ always, regardless of the type of constraints used.

\section{Conclusion} 
\label{sec:conclusion}

In this work, we pointed out a critical flaw in the analysis of QKD protocols using Cascade, that stems from an improper consideration of the classical communication during Cascade. This leads to the computation of secret key rates that are not justified. We proposed a simple and elegant fix,  involving the construction of a convenient virtual protocol that cannot leak less information to Eve than the one using Cascade. Therefore, its key rate can be safely used in any protocol using Cascade. Our approach is easy and straightforward to implement in the numerical framework of \cite{wang2021Numerical}. We applied our solution to various implementations of the BB84 protocol, and compared our results with those of earlier, incorrect approaches. In many cases, we found that the numerical value of the key rate does not change, indicating that the communication from Bob to Alice does not leak additional information to Eve. A number of such cases were shown to arise due to symmetries in the protocol, and in the observed statistics. 
 All code used in this work will be made available soon.

\begin{acknowledgements}

We would like to thank Lars Kamin and Shlok Nahar for giving useful feedback on early drafts of this work. We would like to thank John Burniston and Scott Johnstun for help with debugging code. We would like to thank Louis Salvail for helpful discussions on the Cascade protocol, and for sending us his PhD thesis. This work was funded by the NSERC Discovery Grant, and was conducted at the Institute for Quantum Computing, University of Waterloo, which is funded by Government of Canada through ISED.
\end{acknowledgements}

\bibliography{library_cascade}

\begin{thebibliography}{44}%
\makeatletter
\providecommand \@ifxundefined [1]{%
 \@ifx{#1\undefined}
}%
\providecommand \@ifnum [1]{%
 \ifnum #1\expandafter \@firstoftwo
 \else \expandafter \@secondoftwo
 \fi
}%
\providecommand \@ifx [1]{%
 \ifx #1\expandafter \@firstoftwo
 \else \expandafter \@secondoftwo
 \fi
}%
\providecommand \natexlab [1]{#1}%
\providecommand \enquote  [1]{``#1''}%
\providecommand \bibnamefont  [1]{#1}%
\providecommand \bibfnamefont [1]{#1}%
\providecommand \citenamefont [1]{#1}%
\providecommand \href@noop [0]{\@secondoftwo}%
\providecommand \href [0]{\begingroup \@sanitize@url \@href}%
\providecommand \@href[1]{\@@startlink{#1}\@@href}%
\providecommand \@@href[1]{\endgroup#1\@@endlink}%
\providecommand \@sanitize@url [0]{\catcode `\\12\catcode `\$12\catcode
  `\&12\catcode `\#12\catcode `\^12\catcode `\_12\catcode `\%12\relax}%
\providecommand \@@startlink[1]{}%
\providecommand \@@endlink[0]{}%
\providecommand \url  [0]{\begingroup\@sanitize@url \@url }%
\providecommand \@url [1]{\endgroup\@href {#1}{\urlprefix }}%
\providecommand \urlprefix  [0]{URL }%
\providecommand \Eprint [0]{\href }%
\providecommand \doibase [0]{https://doi.org/}%
\providecommand \selectlanguage [0]{\@gobble}%
\providecommand \bibinfo  [0]{\@secondoftwo}%
\providecommand \bibfield  [0]{\@secondoftwo}%
\providecommand \translation [1]{[#1]}%
\providecommand \BibitemOpen [0]{}%
\providecommand \bibitemStop [0]{}%
\providecommand \bibitemNoStop [0]{.\EOS\space}%
\providecommand \EOS [0]{\spacefactor3000\relax}%
\providecommand \BibitemShut  [1]{\csname bibitem#1\endcsname}%
\let\auto@bib@innerbib\@empty
\bibitem [{\citenamefont {Bennett}\ and\ \citenamefont
  {Brassard}(2014)}]{bennett2014Quantum}%
  \BibitemOpen
  \bibfield  {author} {\bibinfo {author} {\bibfnamefont {C.~H.}\ \bibnamefont
  {Bennett}}\ and\ \bibinfo {author} {\bibfnamefont {G.}~\bibnamefont
  {Brassard}},\ }\bibfield  {title} {\bibinfo {title} {Quantum cryptography:
  {{Public}} key distribution and coin tossing},\ }\href
  {https://doi.org/10.1016/j.tcs.2014.05.025} {\bibfield  {journal} {\bibinfo
  {journal} {Theoretical Computer Science}\ }\textbf {\bibinfo {volume}
  {560}},\ \bibinfo {pages} {7} (\bibinfo {year} {2014})},\ \Eprint
  {https://arxiv.org/abs/2003.06557} {arxiv:2003.06557 [quant-ph]} \BibitemShut
  {NoStop}%
\bibitem [{\citenamefont {Bruss}(1998)}]{bruss1998Optimal}%
  \BibitemOpen
  \bibfield  {author} {\bibinfo {author} {\bibfnamefont {D.}~\bibnamefont
  {Bruss}},\ }\bibfield  {title} {\bibinfo {title} {Optimal eavesdropping in
  quantum cryptography with six states},\ }\href
  {https://doi.org/10.1103/PhysRevLett.81.3018} {\bibfield  {journal} {\bibinfo
   {journal} {Physical Review Letters}\ }\textbf {\bibinfo {volume} {81}},\
  \bibinfo {pages} {3018} (\bibinfo {year} {1998})},\ \Eprint
  {https://arxiv.org/abs/quant-ph/9805019} {arxiv:quant-ph/9805019}
  \BibitemShut {NoStop}%
\bibitem [{\citenamefont {Bennett}\ \emph {et~al.}(1992)\citenamefont
  {Bennett}, \citenamefont {Brassard},\ and\ \citenamefont
  {Mermin}}]{bennett1992Quantum}%
  \BibitemOpen
  \bibfield  {author} {\bibinfo {author} {\bibfnamefont {C.~H.}\ \bibnamefont
  {Bennett}}, \bibinfo {author} {\bibfnamefont {G.}~\bibnamefont {Brassard}},\
  and\ \bibinfo {author} {\bibfnamefont {N.~D.}\ \bibnamefont {Mermin}},\
  }\bibfield  {title} {\bibinfo {title} {Quantum cryptography without
  {{Bell}}'s theorem},\ }\href {https://doi.org/10.1103/PhysRevLett.68.557}
  {\bibfield  {journal} {\bibinfo  {journal} {Physical Review Letters}\
  }\textbf {\bibinfo {volume} {68}},\ \bibinfo {pages} {557} (\bibinfo {year}
  {1992})}\BibitemShut {NoStop}%
\bibitem [{\citenamefont {Brassard}\ and\ \citenamefont
  {Salvail}(1994)}]{brassard1994Secretkey}%
  \BibitemOpen
  \bibfield  {author} {\bibinfo {author} {\bibfnamefont {G.}~\bibnamefont
  {Brassard}}\ and\ \bibinfo {author} {\bibfnamefont {L.}~\bibnamefont
  {Salvail}},\ }\bibfield  {title} {\bibinfo {title} {Secret-key reconciliation
  by public discussion},\ }\href {https://doi.org/10.1007/3-540-48285-7_35}
  {\bibfield  {journal} {\bibinfo  {journal} {Lecture Notes in Computer Science
  (including subseries Lecture Notes in Artificial Intelligence and Lecture
  Notes in Bioinformatics)}\ }\textbf {\bibinfo {volume} {765 LNCS}},\ \bibinfo
  {pages} {410} (\bibinfo {year} {1994})}\BibitemShut {NoStop}%
\bibitem [{\citenamefont {Reis}(2019)}]{reis2019Quantum}%
  \BibitemOpen
  \bibfield  {author} {\bibinfo {author} {\bibfnamefont {A.}~\bibnamefont
  {Reis}},\ }\bibfield  {title} {\bibinfo {title} {Quantum {{Key Distribution
  Post Processing}} - {{A}} study on the {{Information Reconciliation Cascade
  Protocol}}},\ }\href@noop {} {\  (\bibinfo {year} {2019})}\BibitemShut
  {NoStop}%
\bibitem [{\citenamefont {Pedersen}\ and\ \citenamefont
  {Toyran}(2015)}]{pedersen2015High}%
  \BibitemOpen
  \bibfield  {author} {\bibinfo {author} {\bibfnamefont {T.~B.}\ \bibnamefont
  {Pedersen}}\ and\ \bibinfo {author} {\bibfnamefont {M.}~\bibnamefont
  {Toyran}},\ }\bibfield  {title} {\bibinfo {title} {High performance
  information reconciliation for {{QKD}} with {{CASCADE}}},\ }\href
  {https://doi.org/10.26421/qic15.5-6-4} {\bibfield  {journal} {\bibinfo
  {journal} {Quantum Information and Computation}\ ,\ \bibinfo {pages} {419}}
  (\bibinfo {year} {2015})},\ \Eprint {https://arxiv.org/abs/1307.7829}
  {arxiv:1307.7829} \BibitemShut {NoStop}%
\bibitem [{\citenamefont {{Martinez-Mateo}}\ \emph {et~al.}(2015)\citenamefont
  {{Martinez-Mateo}}, \citenamefont {Pacher}, \citenamefont {Peev},
  \citenamefont {Ciurana},\ and\ \citenamefont
  {Martin}}]{martinez-mateo2015Demystifying}%
  \BibitemOpen
  \bibfield  {author} {\bibinfo {author} {\bibfnamefont {J.}~\bibnamefont
  {{Martinez-Mateo}}}, \bibinfo {author} {\bibfnamefont {C.}~\bibnamefont
  {Pacher}}, \bibinfo {author} {\bibfnamefont {M.}~\bibnamefont {Peev}},
  \bibinfo {author} {\bibfnamefont {A.}~\bibnamefont {Ciurana}},\ and\ \bibinfo
  {author} {\bibfnamefont {V.}~\bibnamefont {Martin}},\ }\bibfield  {title}
  {\bibinfo {title} {Demystifying the information reconciliation protocol
  cascade},\ }\href {https://doi.org/10.26421/qic15.5-6-6} {\bibfield
  {journal} {\bibinfo  {journal} {Quantum Information and Computation}\
  }\textbf {\bibinfo {volume} {15}},\ \bibinfo {pages} {453} (\bibinfo {year}
  {2015})},\ \Eprint {https://arxiv.org/abs/1407.3257} {arxiv:1407.3257}
  \BibitemShut {NoStop}%
\bibitem [{\citenamefont {Elkouss}\ \emph {et~al.}(2009)\citenamefont
  {Elkouss}, \citenamefont {Leverrier}, \citenamefont {Al{\'l}eaume},\ and\
  \citenamefont {Boutros}}]{elkouss2009Efficient}%
  \BibitemOpen
  \bibfield  {author} {\bibinfo {author} {\bibfnamefont {D.}~\bibnamefont
  {Elkouss}}, \bibinfo {author} {\bibfnamefont {A.}~\bibnamefont {Leverrier}},
  \bibinfo {author} {\bibfnamefont {R.}~\bibnamefont {Al{\'l}eaume}},\ and\
  \bibinfo {author} {\bibfnamefont {J.~J.}\ \bibnamefont {Boutros}},\
  }\bibfield  {title} {\bibinfo {title} {Efficient reconciliation protocol for
  discrete-variable quantum key distribution},\ }\href
  {https://doi.org/10.1109/ISIT.2009.5205475} {\bibfield  {journal} {\bibinfo
  {journal} {IEEE International Symposium on Information Theory - Proceedings}\
  ,\ \bibinfo {pages} {1879}} (\bibinfo {year} {2009})},\ \Eprint
  {https://arxiv.org/abs/0901.2140} {arxiv:0901.2140} \BibitemShut {NoStop}%
\bibitem [{\citenamefont {Calver}\ \emph {et~al.}(2011)\citenamefont {Calver},
  \citenamefont {Grimaila},\ and\ \citenamefont
  {Humphries}}]{calver2011empirical}%
  \BibitemOpen
  \bibfield  {author} {\bibinfo {author} {\bibfnamefont {T.}~\bibnamefont
  {Calver}}, \bibinfo {author} {\bibfnamefont {M.}~\bibnamefont {Grimaila}},\
  and\ \bibinfo {author} {\bibfnamefont {J.}~\bibnamefont {Humphries}},\
  }\bibfield  {title} {\bibinfo {title} {An empirical analysis of the cascade
  error reconciliation protocol for quantum key distribution},\ }\href
  {https://doi.org/10.1145/2179298.2179363} {\bibfield  {journal} {\bibinfo
  {journal} {ACM International Conference Proceeding Series}\ ,\ \bibinfo
  {pages} {11}} (\bibinfo {year} {2011})}\BibitemShut {NoStop}%
\bibitem [{\citenamefont {Mao}\ \emph {et~al.}(2021)\citenamefont {Mao},
  \citenamefont {Li},\ and\ \citenamefont {Hao}}]{mao2021High}%
  \BibitemOpen
  \bibfield  {author} {\bibinfo {author} {\bibfnamefont {H.-k.}\ \bibnamefont
  {Mao}}, \bibinfo {author} {\bibfnamefont {Q.}~\bibnamefont {Li}},\ and\
  \bibinfo {author} {\bibfnamefont {P.-l.}\ \bibnamefont {Hao}},\ }\bibfield
  {title} {\bibinfo {title} {High performance reconciliation for practical
  quantum key distribution systems},\ }\href@noop {} {\  (\bibinfo {year}
  {2021})},\ \Eprint {https://arxiv.org/abs/2101.12565v2} {arxiv:2101.12565v2}
  \BibitemShut {NoStop}%
\bibitem [{\citenamefont {Pacher}\ \emph {et~al.}(2015)\citenamefont {Pacher},
  \citenamefont {Grabenweger}, \citenamefont {{Martinez-Mateo}},\ and\
  \citenamefont {Martin}}]{pacher2015information}%
  \BibitemOpen
  \bibfield  {author} {\bibinfo {author} {\bibfnamefont {C.}~\bibnamefont
  {Pacher}}, \bibinfo {author} {\bibfnamefont {P.}~\bibnamefont {Grabenweger}},
  \bibinfo {author} {\bibfnamefont {J.}~\bibnamefont {{Martinez-Mateo}}},\ and\
  \bibinfo {author} {\bibfnamefont {V.}~\bibnamefont {Martin}},\ }\bibfield
  {title} {\bibinfo {title} {An information reconciliation protocol for
  secret-key agreement with small leakage},\ }in\ \href
  {https://doi.org/10.1109/ISIT.2015.7282551} {\emph {\bibinfo {booktitle}
  {2015 {{IEEE International Symposium}} on {{Information Theory}}
  ({{ISIT}})}}}\ (\bibinfo {year} {2015})\ pp.\ \bibinfo {pages}
  {730--734}\BibitemShut {NoStop}%
\bibitem [{\citenamefont {Erven}\ \emph {et~al.}(2008)\citenamefont {Erven},
  \citenamefont {Couteau}, \citenamefont {Laflamme},\ and\ \citenamefont
  {Weihs}}]{erven2008Entangled}%
  \BibitemOpen
  \bibfield  {author} {\bibinfo {author} {\bibfnamefont {C.}~\bibnamefont
  {Erven}}, \bibinfo {author} {\bibfnamefont {C.}~\bibnamefont {Couteau}},
  \bibinfo {author} {\bibfnamefont {R.}~\bibnamefont {Laflamme}},\ and\
  \bibinfo {author} {\bibfnamefont {G.}~\bibnamefont {Weihs}},\ }\bibfield
  {title} {\bibinfo {title} {Entangled {{Quantum Key Distribution Over Two
  Free-Space Optical Links}}},\ }\href {https://doi.org/10.1364/OE.16.016840}
  {\bibfield  {journal} {\bibinfo  {journal} {Optics Express}\ }\textbf
  {\bibinfo {volume} {16}},\ \bibinfo {pages} {16840} (\bibinfo {year}
  {2008})},\ \Eprint {https://arxiv.org/abs/0807.2289} {arxiv:0807.2289
  [quant-ph]} \BibitemShut {NoStop}%
\bibitem [{\citenamefont {Dixon}\ \emph {et~al.}(2017)\citenamefont {Dixon},
  \citenamefont {Dynes}, \citenamefont {Lucamarini}, \citenamefont
  {Fr{\"o}hlich}, \citenamefont {Sharpe}, \citenamefont {Plews}, \citenamefont
  {Tam}, \citenamefont {Yuan}, \citenamefont {Tanizawa}, \citenamefont {Sato},
  \citenamefont {Kawamura}, \citenamefont {Fujiwara}, \citenamefont {Sasaki},\
  and\ \citenamefont {Shields}}]{dixon2017Quantum}%
  \BibitemOpen
  \bibfield  {author} {\bibinfo {author} {\bibfnamefont {A.~R.}\ \bibnamefont
  {Dixon}}, \bibinfo {author} {\bibfnamefont {J.~F.}\ \bibnamefont {Dynes}},
  \bibinfo {author} {\bibfnamefont {M.}~\bibnamefont {Lucamarini}}, \bibinfo
  {author} {\bibfnamefont {B.}~\bibnamefont {Fr{\"o}hlich}}, \bibinfo {author}
  {\bibfnamefont {A.~W.}\ \bibnamefont {Sharpe}}, \bibinfo {author}
  {\bibfnamefont {A.}~\bibnamefont {Plews}}, \bibinfo {author} {\bibfnamefont
  {W.}~\bibnamefont {Tam}}, \bibinfo {author} {\bibfnamefont {Z.~L.}\
  \bibnamefont {Yuan}}, \bibinfo {author} {\bibfnamefont {Y.}~\bibnamefont
  {Tanizawa}}, \bibinfo {author} {\bibfnamefont {H.}~\bibnamefont {Sato}},
  \bibinfo {author} {\bibfnamefont {S.}~\bibnamefont {Kawamura}}, \bibinfo
  {author} {\bibfnamefont {M.}~\bibnamefont {Fujiwara}}, \bibinfo {author}
  {\bibfnamefont {M.}~\bibnamefont {Sasaki}},\ and\ \bibinfo {author}
  {\bibfnamefont {A.~J.}\ \bibnamefont {Shields}},\ }\bibfield  {title}
  {\bibinfo {title} {Quantum key distribution with hacking countermeasures and
  long term field trial},\ }\href {https://doi.org/10.1038/s41598-017-01884-0}
  {\bibfield  {journal} {\bibinfo  {journal} {Scientific Reports}\ }\textbf
  {\bibinfo {volume} {7}},\ \bibinfo {pages} {1978} (\bibinfo {year}
  {2017})}\BibitemShut {NoStop}%
\bibitem [{\citenamefont {Su}\ \emph {et~al.}(2009)\citenamefont {Su},
  \citenamefont {Wang}, \citenamefont {Wang}, \citenamefont {Jia},
  \citenamefont {Xie},\ and\ \citenamefont {Peng}}]{su2009Continuous}%
  \BibitemOpen
  \bibfield  {author} {\bibinfo {author} {\bibfnamefont {X.}~\bibnamefont
  {Su}}, \bibinfo {author} {\bibfnamefont {W.}~\bibnamefont {Wang}}, \bibinfo
  {author} {\bibfnamefont {Y.}~\bibnamefont {Wang}}, \bibinfo {author}
  {\bibfnamefont {X.}~\bibnamefont {Jia}}, \bibinfo {author} {\bibfnamefont
  {C.}~\bibnamefont {Xie}},\ and\ \bibinfo {author} {\bibfnamefont
  {K.}~\bibnamefont {Peng}},\ }\bibfield  {title} {\bibinfo {title} {Continuous
  variable quantum key distribution based on optical entangled states without
  signal modulation},\ }\href {https://doi.org/10.1209/0295-5075/87/20005}
  {\bibfield  {journal} {\bibinfo  {journal} {EPL (Europhysics Letters)}\
  }\textbf {\bibinfo {volume} {87}},\ \bibinfo {pages} {20005} (\bibinfo {year}
  {2009})}\BibitemShut {NoStop}%
\bibitem [{\citenamefont {Gobby}\ \emph {et~al.}(2004)\citenamefont {Gobby},
  \citenamefont {Yuan},\ and\ \citenamefont {Shields}}]{gobby2004Quantum}%
  \BibitemOpen
  \bibfield  {author} {\bibinfo {author} {\bibfnamefont {C.}~\bibnamefont
  {Gobby}}, \bibinfo {author} {\bibfnamefont {Z.~L.}\ \bibnamefont {Yuan}},\
  and\ \bibinfo {author} {\bibfnamefont {A.~J.}\ \bibnamefont {Shields}},\
  }\bibfield  {title} {\bibinfo {title} {Quantum key distribution over 122 km
  of standard telecom fiber},\ }\href {https://doi.org/10.1063/1.1738173}
  {\bibfield  {journal} {\bibinfo  {journal} {Applied Physics Letters}\
  }\textbf {\bibinfo {volume} {84}},\ \bibinfo {pages} {3762} (\bibinfo {year}
  {2004})}\BibitemShut {NoStop}%
\bibitem [{\citenamefont {Tentrup}\ \emph {et~al.}(2019)\citenamefont
  {Tentrup}, \citenamefont {Luiten}, \citenamefont {van~der Meer},
  \citenamefont {Hooijschuur},\ and\ \citenamefont
  {Pinkse}}]{tentrup2019Largealphabet}%
  \BibitemOpen
  \bibfield  {author} {\bibinfo {author} {\bibfnamefont {T.~B.~H.}\
  \bibnamefont {Tentrup}}, \bibinfo {author} {\bibfnamefont {W.~M.}\
  \bibnamefont {Luiten}}, \bibinfo {author} {\bibfnamefont {R.}~\bibnamefont
  {van~der Meer}}, \bibinfo {author} {\bibfnamefont {P.}~\bibnamefont
  {Hooijschuur}},\ and\ \bibinfo {author} {\bibfnamefont {P.~W.~H.}\
  \bibnamefont {Pinkse}},\ }\bibfield  {title} {\bibinfo {title}
  {Large-alphabet quantum key distribution using spatially encoded light},\
  }\href {https://doi.org/10.1088/1367-2630/ab5cbe} {\bibfield  {journal}
  {\bibinfo  {journal} {New Journal of Physics}\ }\textbf {\bibinfo {volume}
  {21}},\ \bibinfo {pages} {123044} (\bibinfo {year} {2019})}\BibitemShut
  {NoStop}%
\bibitem [{\citenamefont {Lorenz}\ \emph {et~al.}(2004)\citenamefont {Lorenz},
  \citenamefont {Korolkova},\ and\ \citenamefont
  {Leuchs}}]{lorenz2004Continuousvariable}%
  \BibitemOpen
  \bibfield  {author} {\bibinfo {author} {\bibfnamefont {S.}~\bibnamefont
  {Lorenz}}, \bibinfo {author} {\bibfnamefont {N.}~\bibnamefont {Korolkova}},\
  and\ \bibinfo {author} {\bibfnamefont {G.}~\bibnamefont {Leuchs}},\
  }\bibfield  {title} {\bibinfo {title} {Continuous-variable quantum key
  distribution using polarization encoding and post selection},\ }\href
  {https://doi.org/10.1007/s00340-004-1574-7} {\bibfield  {journal} {\bibinfo
  {journal} {Applied Physics B}\ }\textbf {\bibinfo {volume} {79}},\ \bibinfo
  {pages} {273} (\bibinfo {year} {2004})}\BibitemShut {NoStop}%
\bibitem [{\citenamefont {Winick}\ \emph {et~al.}(2018)\citenamefont {Winick},
  \citenamefont {L{\"u}tkenhaus},\ and\ \citenamefont {Coles}}]{Winick2018}%
  \BibitemOpen
  \bibfield  {author} {\bibinfo {author} {\bibfnamefont {A.}~\bibnamefont
  {Winick}}, \bibinfo {author} {\bibfnamefont {N.}~\bibnamefont
  {L{\"u}tkenhaus}},\ and\ \bibinfo {author} {\bibfnamefont {P.~J.}\
  \bibnamefont {Coles}},\ }\bibfield  {title} {\bibinfo {title} {Reliable
  numerical key rates for quantum key distribution},\ }\href
  {https://doi.org/10.22331/q-2018-07-26-77} {\bibfield  {journal} {\bibinfo
  {journal} {Quantum}\ }\textbf {\bibinfo {volume} {2}},\ \bibinfo {pages} {77}
  (\bibinfo {year} {2018})},\ \Eprint {https://arxiv.org/abs/1710.05511}
  {arxiv:1710.05511} \BibitemShut {NoStop}%
\bibitem [{\citenamefont {Renner}(2016)}]{Renner2016}%
  \BibitemOpen
  \bibfield  {author} {\bibinfo {author} {\bibfnamefont {R.}~\bibnamefont
  {Renner}},\ }\bibfield  {title} {\bibinfo {title} {Security of {{Quantum Key
  Distribution}}},\ }\href {https://doi.org/10.1109/ACCESS.2016.2528227}
  {\bibfield  {journal} {\bibinfo  {journal} {IEEE Access}\ }\textbf {\bibinfo
  {volume} {4}},\ \bibinfo {pages} {724} (\bibinfo {year} {2016})},\ \Eprint
  {https://arxiv.org/abs/quant-ph/0512258} {arxiv:quant-ph/0512258}
  \BibitemShut {NoStop}%
\bibitem [{\citenamefont {Christandl}\ \emph {et~al.}(2009)\citenamefont
  {Christandl}, \citenamefont {K{\"o}nig},\ and\ \citenamefont
  {Renner}}]{christandl2009Postselection}%
  \BibitemOpen
  \bibfield  {author} {\bibinfo {author} {\bibfnamefont {M.}~\bibnamefont
  {Christandl}}, \bibinfo {author} {\bibfnamefont {R.}~\bibnamefont
  {K{\"o}nig}},\ and\ \bibinfo {author} {\bibfnamefont {R.}~\bibnamefont
  {Renner}},\ }\bibfield  {title} {\bibinfo {title} {Postselection technique
  for quantum channels with applications to quantum cryptography},\ }\href
  {https://doi.org/10.1103/PhysRevLett.102.020504} {\bibfield  {journal}
  {\bibinfo  {journal} {Physical Review Letters}\ }\textbf {\bibinfo {volume}
  {102}},\ \bibinfo {pages} {1} (\bibinfo {year} {2009})},\ \Eprint
  {https://arxiv.org/abs/0809.3019} {arxiv:0809.3019} \BibitemShut {NoStop}%
\bibitem [{\citenamefont {George}\ \emph {et~al.}(2021)\citenamefont {George},
  \citenamefont {Lin},\ and\ \citenamefont {L{\"u}tkenhaus}}]{George2021}%
  \BibitemOpen
  \bibfield  {author} {\bibinfo {author} {\bibfnamefont {I.}~\bibnamefont
  {George}}, \bibinfo {author} {\bibfnamefont {J.}~\bibnamefont {Lin}},\ and\
  \bibinfo {author} {\bibfnamefont {N.}~\bibnamefont {L{\"u}tkenhaus}},\
  }\bibfield  {title} {\bibinfo {title} {Numerical calculations of the finite
  key rate for general quantum key distribution protocols},\ }\bibfield
  {journal} {\bibinfo  {journal} {Physical Review Research}\ }\textbf {\bibinfo
  {volume} {3}},\ \href {https://doi.org/10.1103/PhysRevResearch.3.013274}
  {10.1103/PhysRevResearch.3.013274} (\bibinfo {year} {2021}),\ \Eprint
  {https://arxiv.org/abs/2004.11865} {arxiv:2004.11865} \BibitemShut {NoStop}%
\bibitem [{\citenamefont {Scarani}\ and\ \citenamefont
  {Renner}(2008{\natexlab{a}})}]{valerio_2008}%
  \BibitemOpen
  \bibfield  {author} {\bibinfo {author} {\bibfnamefont {V.}~\bibnamefont
  {Scarani}}\ and\ \bibinfo {author} {\bibfnamefont {R.}~\bibnamefont
  {Renner}},\ }\bibfield  {title} {\bibinfo {title} {Quantum cryptography with
  finite resources: Unconditional security bound for discrete-variable
  protocols with one-way postprocessing},\ }\href
  {https://doi.org/10.1103/PhysRevLett.100.200501} {\bibfield  {journal}
  {\bibinfo  {journal} {Phys. Rev. Lett.}\ }\textbf {\bibinfo {volume} {100}},\
  \bibinfo {pages} {200501} (\bibinfo {year} {2008}{\natexlab{a}})}\BibitemShut
  {NoStop}%
\bibitem [{\citenamefont {Scarani}\ and\ \citenamefont
  {Renner}(2008{\natexlab{b}})}]{valeriotwoway}%
  \BibitemOpen
  \bibfield  {author} {\bibinfo {author} {\bibfnamefont {V.}~\bibnamefont
  {Scarani}}\ and\ \bibinfo {author} {\bibfnamefont {R.}~\bibnamefont
  {Renner}},\ }\bibfield  {title} {\bibinfo {title} {Security bounds for
  quantum cryptography with finite resources},\ }in\ \href@noop {} {\emph
  {\bibinfo {booktitle} {Theory of Quantum Computation, Communication, and
  Cryptography}}},\ \bibinfo {editor} {edited by\ \bibinfo {editor}
  {\bibfnamefont {Y.}~\bibnamefont {Kawano}}\ and\ \bibinfo {editor}
  {\bibfnamefont {M.}~\bibnamefont {Mosca}}}\ (\bibinfo  {publisher} {Springer
  Berlin Heidelberg},\ \bibinfo {address} {Berlin, Heidelberg},\ \bibinfo
  {year} {2008})\ pp.\ \bibinfo {pages} {83--95}\BibitemShut {NoStop}%
\bibitem [{\citenamefont {Tomamichel}\ and\ \citenamefont
  {Leverrier}(2017)}]{Tomamichel_2017}%
  \BibitemOpen
  \bibfield  {author} {\bibinfo {author} {\bibfnamefont {M.}~\bibnamefont
  {Tomamichel}}\ and\ \bibinfo {author} {\bibfnamefont {A.}~\bibnamefont
  {Leverrier}},\ }\bibfield  {title} {\bibinfo {title} {A largely
  self-contained and complete security proof for quantum key distribution},\
  }\href {https://doi.org/10.22331/q-2017-07-14-14} {\bibfield  {journal}
  {\bibinfo  {journal} {Quantum}\ }\textbf {\bibinfo {volume} {1}},\ \bibinfo
  {pages} {14} (\bibinfo {year} {2017})}\BibitemShut {NoStop}%
\bibitem [{\citenamefont {Curty}\ \emph {et~al.}(2004)\citenamefont {Curty},
  \citenamefont {Lewenstein},\ and\ \citenamefont
  {L{\"u}tkenhaus}}]{curty2004Entanglement}%
  \BibitemOpen
  \bibfield  {author} {\bibinfo {author} {\bibfnamefont {M.}~\bibnamefont
  {Curty}}, \bibinfo {author} {\bibfnamefont {M.}~\bibnamefont {Lewenstein}},\
  and\ \bibinfo {author} {\bibfnamefont {N.}~\bibnamefont {L{\"u}tkenhaus}},\
  }\bibfield  {title} {\bibinfo {title} {Entanglement as a {{Precondition}} for
  {{Secure Quantum Key Distribution}}},\ }\href
  {https://doi.org/10.1103/PhysRevLett.92.217903} {\bibfield  {journal}
  {\bibinfo  {journal} {Physical Review Letters}\ }\textbf {\bibinfo {volume}
  {92}},\ \bibinfo {pages} {217903} (\bibinfo {year} {2004})}\BibitemShut
  {NoStop}%
\bibitem [{\citenamefont {Wang}\ and\ \citenamefont
  {L\"utkenhaus}(2022)}]{wang2021Numerical}%
  \BibitemOpen
  \bibfield  {author} {\bibinfo {author} {\bibfnamefont {W.}~\bibnamefont
  {Wang}}\ and\ \bibinfo {author} {\bibfnamefont {N.}~\bibnamefont
  {L\"utkenhaus}},\ }\bibfield  {title} {\bibinfo {title} {Numerical security
  proof for the decoy-state bb84 protocol and measurement-device-independent
  quantum key distribution resistant against large basis misalignment},\ }\href
  {https://doi.org/10.1103/PhysRevResearch.4.043097} {\bibfield  {journal}
  {\bibinfo  {journal} {Phys. Rev. Res.}\ }\textbf {\bibinfo {volume} {4}},\
  \bibinfo {pages} {043097} (\bibinfo {year} {2022})}\BibitemShut {NoStop}%
\bibitem [{\citenamefont {Devetak}\ and\ \citenamefont
  {Winter}(2005)}]{Devetak2005}%
  \BibitemOpen
  \bibfield  {author} {\bibinfo {author} {\bibfnamefont {I.}~\bibnamefont
  {Devetak}}\ and\ \bibinfo {author} {\bibfnamefont {A.}~\bibnamefont
  {Winter}},\ }\bibfield  {title} {\bibinfo {title} {Distillation of secret key
  and entanglement from quantum states},\ }\href
  {https://doi.org/10.1098/rspa.2004.1372} {\bibfield  {journal} {\bibinfo
  {journal} {Proceedings of the Royal Society A: Mathematical, Physical and
  Engineering Sciences}\ }\textbf {\bibinfo {volume} {461}},\ \bibinfo {pages}
  {207} (\bibinfo {year} {2005})},\ \Eprint
  {https://arxiv.org/abs/quant-ph/0306078} {arxiv:quant-ph/0306078}
  \BibitemShut {NoStop}%
\bibitem [{\citenamefont {Xu}\ \emph {et~al.}(2020)\citenamefont {Xu},
  \citenamefont {Ma}, \citenamefont {Zhang}, \citenamefont {Lo},\ and\
  \citenamefont {Pan}}]{xu2020Secure}%
  \BibitemOpen
  \bibfield  {author} {\bibinfo {author} {\bibfnamefont {F.}~\bibnamefont
  {Xu}}, \bibinfo {author} {\bibfnamefont {X.}~\bibnamefont {Ma}}, \bibinfo
  {author} {\bibfnamefont {Q.}~\bibnamefont {Zhang}}, \bibinfo {author}
  {\bibfnamefont {H.-K.}\ \bibnamefont {Lo}},\ and\ \bibinfo {author}
  {\bibfnamefont {J.-W.}\ \bibnamefont {Pan}},\ }\bibfield  {title} {\bibinfo
  {title} {Secure quantum key distribution with realistic devices},\ }\href
  {https://doi.org/10.1103/RevModPhys.92.025002} {\bibfield  {journal}
  {\bibinfo  {journal} {Reviews of Modern Physics}\ }\textbf {\bibinfo {volume}
  {92}},\ \bibinfo {pages} {025002} (\bibinfo {year} {2020})}\BibitemShut
  {NoStop}%
\bibitem [{\citenamefont {Scarani}\ \emph {et~al.}(2009)\citenamefont
  {Scarani}, \citenamefont {{Bechmann-Pasquinucci}}, \citenamefont {Cerf},
  \citenamefont {Du{\v s}ek}, \citenamefont {L{\"u}tkenhaus},\ and\
  \citenamefont {Peev}}]{Scarani2009}%
  \BibitemOpen
  \bibfield  {author} {\bibinfo {author} {\bibfnamefont {V.}~\bibnamefont
  {Scarani}}, \bibinfo {author} {\bibfnamefont {H.}~\bibnamefont
  {{Bechmann-Pasquinucci}}}, \bibinfo {author} {\bibfnamefont {N.~J.}\
  \bibnamefont {Cerf}}, \bibinfo {author} {\bibfnamefont {M.}~\bibnamefont
  {Du{\v s}ek}}, \bibinfo {author} {\bibfnamefont {N.}~\bibnamefont
  {L{\"u}tkenhaus}},\ and\ \bibinfo {author} {\bibfnamefont {M.}~\bibnamefont
  {Peev}},\ }\bibfield  {title} {\bibinfo {title} {The security of practical
  quantum key distribution},\ }\href
  {https://doi.org/10.1103/RevModPhys.81.1301} {\bibfield  {journal} {\bibinfo
  {journal} {Reviews of Modern Physics}\ }\textbf {\bibinfo {volume} {81}},\
  \bibinfo {pages} {1301} (\bibinfo {year} {2009})},\ \Eprint
  {https://arxiv.org/abs/0802.4155} {arxiv:0802.4155} \BibitemShut {NoStop}%
\bibitem [{\citenamefont {Ferenczi}\ and\ \citenamefont
  {L\"utkenhaus}(2012)}]{ferenczi2011Symmetries}%
  \BibitemOpen
  \bibfield  {author} {\bibinfo {author} {\bibfnamefont {A.}~\bibnamefont
  {Ferenczi}}\ and\ \bibinfo {author} {\bibfnamefont {N.}~\bibnamefont
  {L\"utkenhaus}},\ }\bibfield  {title} {\bibinfo {title} {Symmetries in
  quantum key distribution and the connection between optimal attacks and
  optimal cloning},\ }\href {https://doi.org/10.1103/PhysRevA.85.052310}
  {\bibfield  {journal} {\bibinfo  {journal} {Phys. Rev. A}\ }\textbf {\bibinfo
  {volume} {85}},\ \bibinfo {pages} {052310} (\bibinfo {year}
  {2012})}\BibitemShut {NoStop}%
\bibitem [{\citenamefont {Watanabe}\ \emph {et~al.}(2007)\citenamefont
  {Watanabe}, \citenamefont {Matsumoto}, \citenamefont {Uyematsu},\ and\
  \citenamefont {Kawano}}]{Watanabe2007}%
  \BibitemOpen
  \bibfield  {author} {\bibinfo {author} {\bibfnamefont {S.}~\bibnamefont
  {Watanabe}}, \bibinfo {author} {\bibfnamefont {R.}~\bibnamefont {Matsumoto}},
  \bibinfo {author} {\bibfnamefont {T.}~\bibnamefont {Uyematsu}},\ and\
  \bibinfo {author} {\bibfnamefont {Y.}~\bibnamefont {Kawano}},\ }\bibfield
  {title} {\bibinfo {title} {Key rate of quantum key distribution with hashed
  two-way classical communication},\ }\bibfield  {journal} {\bibinfo  {journal}
  {Physical Review A - Atomic, Molecular, and Optical Physics}\ }\textbf
  {\bibinfo {volume} {76}},\ \href {https://doi.org/10.1103/PhysRevA.76.032312}
  {10.1103/PhysRevA.76.032312} (\bibinfo {year} {2007}),\ \Eprint
  {https://arxiv.org/abs/0705.2904} {arxiv:0705.2904} \BibitemShut {NoStop}%
\bibitem [{\citenamefont {Tupkary}(2022)}]{tupkary2022Improved}%
  \BibitemOpen
  \bibfield  {author} {\bibinfo {author} {\bibfnamefont {D.}~\bibnamefont
  {Tupkary}},\ }\emph {\bibinfo {title} {Improved {{Keyrates}} for {{Quantum
  Key Distribution}} from Two-Way {{Classical Communication}}}},\ \href
  {https://uwspace.uwaterloo.ca/bitstream/handle/10012/18772/Tupkary_Devashish.pdf?sequence=3}
  {Master's thesis},\ \bibinfo  {school} {University of Waterloo} (\bibinfo
  {year} {2022})\BibitemShut {NoStop}%
\bibitem [{\citenamefont {Bennett}\ \emph {et~al.}(1996)\citenamefont
  {Bennett}, \citenamefont {DiVincenzo}, \citenamefont {Smolin},\ and\
  \citenamefont {Wootters}}]{bennett1996Mixed}%
  \BibitemOpen
  \bibfield  {author} {\bibinfo {author} {\bibfnamefont {C.~H.}\ \bibnamefont
  {Bennett}}, \bibinfo {author} {\bibfnamefont {D.~P.}\ \bibnamefont
  {DiVincenzo}}, \bibinfo {author} {\bibfnamefont {J.~A.}\ \bibnamefont
  {Smolin}},\ and\ \bibinfo {author} {\bibfnamefont {W.~K.}\ \bibnamefont
  {Wootters}},\ }\bibfield  {title} {\bibinfo {title} {Mixed {{State
  Entanglement}} and {{Quantum Error Correction}}},\ }\href
  {https://doi.org/10.1103/PhysRevA.54.3824} {\bibfield  {journal} {\bibinfo
  {journal} {Physical Review A}\ }\textbf {\bibinfo {volume} {54}},\ \bibinfo
  {pages} {3824} (\bibinfo {year} {1996})},\ \Eprint
  {https://arxiv.org/abs/quant-ph/9604024} {arxiv:quant-ph/9604024}
  \BibitemShut {NoStop}%
\bibitem [{\citenamefont {Lim}\ \emph {et~al.}(2014)\citenamefont {Lim},
  \citenamefont {Curty}, \citenamefont {Walenta}, \citenamefont {Xu},\ and\
  \citenamefont {Zbinden}}]{lim2014Concise}%
  \BibitemOpen
  \bibfield  {author} {\bibinfo {author} {\bibfnamefont {C.~C.~W.}\
  \bibnamefont {Lim}}, \bibinfo {author} {\bibfnamefont {M.}~\bibnamefont
  {Curty}}, \bibinfo {author} {\bibfnamefont {N.}~\bibnamefont {Walenta}},
  \bibinfo {author} {\bibfnamefont {F.}~\bibnamefont {Xu}},\ and\ \bibinfo
  {author} {\bibfnamefont {H.}~\bibnamefont {Zbinden}},\ }\bibfield  {title}
  {\bibinfo {title} {Concise security bounds for practical decoy-state quantum
  key distribution},\ }\href {https://doi.org/10.1103/PhysRevA.89.022307}
  {\bibfield  {journal} {\bibinfo  {journal} {Physical Review A}\ }\textbf
  {\bibinfo {volume} {89}},\ \bibinfo {pages} {022307} (\bibinfo {year}
  {2014})}\BibitemShut {NoStop}%
\bibitem [{\citenamefont {Lo}\ \emph {et~al.}(2005)\citenamefont {Lo},
  \citenamefont {Ma},\ and\ \citenamefont {Chen}}]{lo2005Decoy}%
  \BibitemOpen
  \bibfield  {author} {\bibinfo {author} {\bibfnamefont {H.-K.}\ \bibnamefont
  {Lo}}, \bibinfo {author} {\bibfnamefont {X.}~\bibnamefont {Ma}},\ and\
  \bibinfo {author} {\bibfnamefont {K.}~\bibnamefont {Chen}},\ }\bibfield
  {title} {\bibinfo {title} {Decoy {{State Quantum Key Distribution}}},\ }\href
  {https://doi.org/10.1103/PhysRevLett.94.230504} {\bibfield  {journal}
  {\bibinfo  {journal} {Physical Review Letters}\ }\textbf {\bibinfo {volume}
  {94}},\ \bibinfo {pages} {230504} (\bibinfo {year} {2005})}\BibitemShut
  {NoStop}%
\bibitem [{\citenamefont {Rusca}\ \emph {et~al.}(2018)\citenamefont {Rusca},
  \citenamefont {Boaron}, \citenamefont {Gr{\"u}nenfelder}, \citenamefont
  {Martin},\ and\ \citenamefont {Zbinden}}]{rusca2018Finitekey}%
  \BibitemOpen
  \bibfield  {author} {\bibinfo {author} {\bibfnamefont {D.}~\bibnamefont
  {Rusca}}, \bibinfo {author} {\bibfnamefont {A.}~\bibnamefont {Boaron}},
  \bibinfo {author} {\bibfnamefont {F.}~\bibnamefont {Gr{\"u}nenfelder}},
  \bibinfo {author} {\bibfnamefont {A.}~\bibnamefont {Martin}},\ and\ \bibinfo
  {author} {\bibfnamefont {H.}~\bibnamefont {Zbinden}},\ }\bibfield  {title}
  {\bibinfo {title} {Finite-key analysis for the 1-decoy state {{QKD}}
  protocol},\ }\href {https://doi.org/10.1063/1.5023340} {\bibfield  {journal}
  {\bibinfo  {journal} {Applied Physics Letters}\ }\textbf {\bibinfo {volume}
  {112}},\ \bibinfo {pages} {171104} (\bibinfo {year} {2018})}\BibitemShut
  {NoStop}%
\bibitem [{\citenamefont {Rice}\ and\ \citenamefont
  {Harrington}(2009)}]{rice2009Numerical}%
  \BibitemOpen
  \bibfield  {author} {\bibinfo {author} {\bibfnamefont {P.}~\bibnamefont
  {Rice}}\ and\ \bibinfo {author} {\bibfnamefont {J.}~\bibnamefont
  {Harrington}},\ }\href {https://doi.org/10.48550/arXiv.0901.0013} {\bibinfo
  {title} {Numerical analysis of decoy state quantum key distribution
  protocols}} (\bibinfo {year} {2009}),\ \Eprint
  {https://arxiv.org/abs/0901.0013} {arxiv:0901.0013 [quant-ph]} \BibitemShut
  {NoStop}%
\bibitem [{\citenamefont {Hwang}(2003)}]{hwang2003Quantum}%
  \BibitemOpen
  \bibfield  {author} {\bibinfo {author} {\bibfnamefont {W.-Y.}\ \bibnamefont
  {Hwang}},\ }\bibfield  {title} {\bibinfo {title} {Quantum {{Key
  Distribution}} with {{High Loss}}: {{Toward Global Secure Communication}}},\
  }\href {https://doi.org/10.1103/PhysRevLett.91.057901} {\bibfield  {journal}
  {\bibinfo  {journal} {Physical Review Letters}\ }\textbf {\bibinfo {volume}
  {91}},\ \bibinfo {pages} {057901} (\bibinfo {year} {2003})}\BibitemShut
  {NoStop}%
\bibitem [{\citenamefont {Ma}\ \emph {et~al.}(2005)\citenamefont {Ma},
  \citenamefont {Qi}, \citenamefont {Zhao},\ and\ \citenamefont
  {Lo}}]{ma2005Practical}%
  \BibitemOpen
  \bibfield  {author} {\bibinfo {author} {\bibfnamefont {X.}~\bibnamefont
  {Ma}}, \bibinfo {author} {\bibfnamefont {B.}~\bibnamefont {Qi}}, \bibinfo
  {author} {\bibfnamefont {Y.}~\bibnamefont {Zhao}},\ and\ \bibinfo {author}
  {\bibfnamefont {H.-K.}\ \bibnamefont {Lo}},\ }\bibfield  {title} {\bibinfo
  {title} {Practical decoy state for quantum key distribution},\ }\href
  {https://doi.org/10.1103/PhysRevA.72.012326} {\bibfield  {journal} {\bibinfo
  {journal} {Physical Review A}\ }\textbf {\bibinfo {volume} {72}},\ \bibinfo
  {pages} {012326} (\bibinfo {year} {2005})}\BibitemShut {NoStop}%
\bibitem [{\citenamefont {Wang}(2005)}]{Wang_2005}%
  \BibitemOpen
  \bibfield  {author} {\bibinfo {author} {\bibfnamefont {X.-B.}\ \bibnamefont
  {Wang}},\ }\bibfield  {title} {\bibinfo {title} {Beating the
  photon-number-splitting attack in practical quantum cryptography},\ }\href
  {https://doi.org/10.1103/PhysRevLett.94.230503} {\bibfield  {journal}
  {\bibinfo  {journal} {Phys. Rev. Lett.}\ }\textbf {\bibinfo {volume} {94}},\
  \bibinfo {pages} {230503} (\bibinfo {year} {2005})}\BibitemShut {NoStop}%
\bibitem [{\citenamefont {Gittsovich}\ \emph {et~al.}(2014)\citenamefont
  {Gittsovich}, \citenamefont {Beaudry}, \citenamefont {Narasimhachar},
  \citenamefont {Alvarez},\ and\ \citenamefont {Moroder}}]{Gittsovich2014}%
  \BibitemOpen
  \bibfield  {author} {\bibinfo {author} {\bibfnamefont {O.}~\bibnamefont
  {Gittsovich}}, \bibinfo {author} {\bibfnamefont {N.~J.}\ \bibnamefont
  {Beaudry}}, \bibinfo {author} {\bibfnamefont {V.}~\bibnamefont
  {Narasimhachar}}, \bibinfo {author} {\bibfnamefont {R.~R.}\ \bibnamefont
  {Alvarez}},\ and\ \bibinfo {author} {\bibfnamefont {T.}~\bibnamefont
  {Moroder}},\ }\bibfield  {title} {\bibinfo {title} {Squashing model for
  detectors and applications to quantum-key-distribution protocols},\ }\href
  {https://doi.org/10.1103/PhysRevA.89.012325} {\bibfield  {journal} {\bibinfo
  {journal} {Physical Review A}\ }\textbf {\bibinfo {volume} {012325}}
  (\bibinfo {year} {2014})}\BibitemShut {NoStop}%
\bibitem [{\citenamefont {Lin}\ \emph {et~al.}(2019)\citenamefont {Lin},
  \citenamefont {Upadhyaya},\ and\ \citenamefont {L{\"u}tkenhaus}}]{Lin2019}%
  \BibitemOpen
  \bibfield  {author} {\bibinfo {author} {\bibfnamefont {J.}~\bibnamefont
  {Lin}}, \bibinfo {author} {\bibfnamefont {T.}~\bibnamefont {Upadhyaya}},\
  and\ \bibinfo {author} {\bibfnamefont {N.}~\bibnamefont {L{\"u}tkenhaus}},\
  }\bibfield  {title} {\bibinfo {title} {Asymptotic {{Security Analysis}} of
  {{Discrete-Modulated Continuous-Variable Quantum Key Distribution}}},\ }\href
  {https://doi.org/10.1103/PhysRevX.9.041064} {\bibfield  {journal} {\bibinfo
  {journal} {Physical Review X}\ }\textbf {\bibinfo {volume} {91}},\ \bibinfo
  {pages} {41064} (\bibinfo {year} {2019})},\ \Eprint
  {https://arxiv.org/abs/1905.10896} {arxiv:1905.10896} \BibitemShut {NoStop}%
\bibitem [{\citenamefont {Li}\ and\ \citenamefont
  {L{\"u}tkenhaus}(2020)}]{li2020Improving}%
  \BibitemOpen
  \bibfield  {author} {\bibinfo {author} {\bibfnamefont {N.~K.~H.}\
  \bibnamefont {Li}}\ and\ \bibinfo {author} {\bibfnamefont {N.}~\bibnamefont
  {L{\"u}tkenhaus}},\ }\bibfield  {title} {\bibinfo {title} {Improving key
  rates of the unbalanced phase-encoded {{BB84}} protocol using the flag-state
  squashing model},\ }\href {https://doi.org/10.1103/PhysRevResearch.2.043172}
  {\bibfield  {journal} {\bibinfo  {journal} {Physical Review Research}\
  }\textbf {\bibinfo {volume} {2}},\ \bibinfo {pages} {043172} (\bibinfo {year}
  {2020})}\BibitemShut {NoStop}%
\bibitem [{\citenamefont {L{\"u}tkenhaus}\ and\ \citenamefont
  {Jahma}(2002)}]{lutkenhaus2002Quantum}%
  \BibitemOpen
  \bibfield  {author} {\bibinfo {author} {\bibfnamefont {N.}~\bibnamefont
  {L{\"u}tkenhaus}}\ and\ \bibinfo {author} {\bibfnamefont {M.}~\bibnamefont
  {Jahma}},\ }\bibfield  {title} {\bibinfo {title} {Quantum key distribution
  with realistic states: Photon-number statistics in the photon-number
  splitting attack},\ }\href {https://doi.org/10.1088/1367-2630/4/1/344}
  {\bibfield  {journal} {\bibinfo  {journal} {New Journal of Physics}\ }\textbf
  {\bibinfo {volume} {4}} (\bibinfo {year} {2002})},\ \Eprint
  {https://arxiv.org/abs/quant-ph/0112147} {arxiv:quant-ph/0112147}
  \BibitemShut {NoStop}%
\end{thebibliography}%

\appendix

\section{Protocol Descriptions}
\label{appendix:protocol_descriptions}
\subsection{Qubit BB84 }
\label{appendix:qubitBB84}
Using the source-replacement scheme \cite{curty2004Entanglement}, the protocol can be equivalently described as Alice creating the Bell-state $\ket{\psi}_{AA^\prime} = \ket{\phi_+} = \frac{\ket{00}+\ket{11}}{\sqrt{2}}$, and sending $A^\prime$ to Bob. We model misalignment as a rotation of angle $\theta$ about the $Y$ axis on $A^\prime$, with 
\begin{equation}
	\begin{aligned}
	U(\theta) &= I_A \otimes \begin{pmatrix}
		\cos(\theta) & -\sin(\theta) \\
		\sin(\theta) & \cos(\theta),
	\end{pmatrix} \\
\mathcal{E}_{\text{misalign}} (\rho) &= U(\theta) \rho U(\theta)^\dagger.
\end{aligned}
\end{equation}
Depolarization is modelled as a map 
\begin{equation}
	\mathcal{E}_\text{depol} (\rho)= (1-q) (\rho)+q \text{Tr}_{A^\prime} (\rho) \otimes \frac{I_B}{2}  .
\end{equation}
The state on which statistics are computed is given by $\rho_{AB} =  	\mathcal{E}_\text{depol} ( \mathcal{E}_{\text{misalign}}(\ket{\phi_+} \bra{\phi_+}) )$. The entries in Table \ref{table:qubit_constraints} can be computed via $\gamma_i = \text{Tr} (\Gamma_i \rho_{AB})$.

Both Alice and Bob perform measurements on qubit systems, and their POVMs are given by $\{ P_{(Z,0)} = p_z \ket{0} \bra{0}, P_{(Z,1)} = p_z \ket{1}\bra{1}, P_{(X,0)} = p_x \ket{+} \bra{+} , P_{(X,1)} = p_x \ket{-} \bra{-} \}$, with $p_z=p_x=\frac{1}{2}$ . In addition, Alice implements the keymap by simply copying the measurement outcome to the key register. 
From the discussion in Appendix A of \cite{Lin2019}, we can remove certain registers created by the generic form of the Kraus operators in Eq.~\eqref{eq:generic_kraus}. In particular, we do not need to consider the registers that store Alice and Bob's outcome, and we only need one copy of the announcement register.
 In this case, the general form for the Kraus operators from Eq.~\eqref{eq:generic_kraus} now becomes
		\begin{equation} \label{eq:new_generic_kraus}
		\begin{aligned}
			K_{\alpha}& = \sum_x \ket{r(\alpha,\alpha,x) }_Z \otimes \sqrt{ \sum_{y} P^A_{(\alpha,x)} \otimes P^B_{(\alpha,y)} }  \otimes  \ket{\alpha}_{\tilde{A}},
		\end{aligned}
	\end{equation}
while Eq.~\eqref{eq:newkraus} becomes
	\begin{equation} \label{eq:new_generic_newkraus}
	\begin{aligned}
		K^\prime_{\alpha,w}& = \sum_x \ket{r(\alpha,\alpha,x) }_Z \otimes \sqrt{ \sum_{ \substack{y \\ x\oplus y = w}} P^A_{(\alpha,x)} \otimes P^B_{(\alpha,y)} } \\
		& \otimes  \ket{\alpha}_{\tilde{A}} \otimes \ket{w}_W,
	\end{aligned}
\end{equation}
where $\alpha,\beta$ denotes basis choice, and $x,y$ denotes measurement outcomes. Alice and Bob's POVMs are given by $P^A = \{P^A_{(\alpha,x)}\}$, and $P^B = \{P^B_{(\alpha,y)}\}$. Since Alice and Bob throw away all signals that have basis mismatch, the set of operators generating the $\mathcal{G}$ map can be given by $\{ K_{\alpha} \} $, and the set of operators generating $\mathcal{G}^\prime$ is given by $\{K^\prime_{\alpha,w}\}$.
 The $\mathcal{Z}$ map has Kraus operators $\{ Z_i\} $ given by $Z_i = \ket{i} \bra{i}_Z \otimes I_{AB \tilde{A} }$.
Therefore, the final Kraus operators for $F$ are given by
	
\begin{equation}
	\begin{aligned}
		K_Z &= \left[ \begin{pmatrix} 1 \\ 0 \end{pmatrix}_Z  \otimes  \sqrt{p_z} \begin{pmatrix} 1 & \\ & 0 \end{pmatrix}_A + \begin{pmatrix} 0 \\ 1 \end{pmatrix}_Z \otimes  \sqrt{p_z} \begin{pmatrix} 0 &  \\ & 1  \end{pmatrix}_A \right] \\
		& \otimes \sqrt{p_z} \begin{pmatrix} 1 & \\ & 1 \end{pmatrix}_B \otimes \begin{pmatrix} 1 \\ 0 \end{pmatrix}_{\tilde{A}}, \\
	K_X &= \left[ \begin{pmatrix} 1 \\ 0 \end{pmatrix}_Z \!  \otimes \sqrt{\frac{p_x}{2} } \begin{pmatrix} 1 & 1 \\ 1 & 1 \end{pmatrix}_A \! + \begin{pmatrix} 0 \\ 1 \end{pmatrix}_Z \!   \otimes\sqrt{ \frac{p_x}{2} } \begin{pmatrix} 1 & -1 \\ -1 & 1 \end{pmatrix}_A  \! \right] \\
	& \otimes \sqrt{p_x} \begin{pmatrix} 1 & \\ & 1 \end{pmatrix}_B \otimes \begin{pmatrix} 0 \\ 1 \end{pmatrix}_{\tilde{A}}, 
\end{aligned}
\end{equation}

 and
\begin{equation}
	\begin{aligned}
		Z_1 &= \begin{pmatrix} 1 & \\ & 0 \end{pmatrix} \otimes \mathbb{I}_{\dim_A \times\dim_B \times 2}, \\
		Z_2 &= \begin{pmatrix} 0 & \\ & 1 \end{pmatrix} \otimes \mathbb{I}_{\dim_A \times\dim_B \times 2}.
	\end{aligned}
\end{equation}
The operators for $F^\prime$ can be constructed from Eq.~\eqref{eq:new_generic_newkraus}.

\subsection{WCP Decoy BB84}
\label{appendix:decoyprotocol}
Along with source-replacement, we use the squashing model from \cite{Gittsovich2014} to squash Bob's system to three dimensions. Since we only generate key from the single-photon pulses, Alice's POVMs are given by $\{ P_{(Z,0)} = p_z \ket{0} \bra{0}, P_{(Z,1)} = p_z \ket{1}\bra{1}, P_{(X,0)} = p_x \ket{+} \bra{+} , P_{(X,1)} = p_x \ket{-} \bra{-} \}$, with $p_z=p_x=\frac{1}{2}$ . Bobs POVMs are given by 
\begin{equation}
	\begin{aligned}
		P^B_{(Z,0)} &= p_z \begin{pmatrix} 0 & 0 &0 \\ 0& 1 & 0\\ 0&0 & 0 \end{pmatrix} , \quad 	P^B_{(Z,1)} = p_z  \begin{pmatrix} 0 & 0& 0\\ 0& 0 & 0\\ 0&0 & 1 \end{pmatrix}, \\
		P^B_{(X,0)} &= \frac{p_x}{2} \begin{pmatrix} 0&0 & 0  \\  0&1 & 1  \\  0& 1 & 1  \end{pmatrix} , \quad 	P^B_{(X,1)} = \frac{p_x}{2} \begin{pmatrix} 0&0 &0   \\  0&1 & -1  \\ 0 & -1 & 1  \end{pmatrix}, \\
		P^B_{\bot} &= \begin{pmatrix} 1 & 0 & 0 \\ 0 & 0 & 0 \\ 0 & 0 & 0 \end{pmatrix},
	\end{aligned}
\end{equation}
with $p_x=p_z= \frac{1}{2}$.
Here, the first column corresponds to the vacuum subspace, while the second and third column make up the qubit subspace.
 Again, from the discussion in \cite{Lin2019}, we can remove certain registers created by the generic form of the Kraus operators in Eq.~\eqref{eq:generic_kraus}. After removing these registers, the form of the Kraus operators is given by Eq.~\eqref{eq:new_generic_kraus}.
Therefore, the final Kraus operators for $F$ are given by

\begin{equation}
	\begin{aligned}
		K_Z &= \left[ \begin{pmatrix} 1 \\ 0 \end{pmatrix}_Z  \otimes \sqrt{p_z} \begin{pmatrix} 1 & \\ & 0 \end{pmatrix}_A  + \begin{pmatrix} 0 \\ 1 \end{pmatrix}_Z \otimes   \sqrt{p_z} \begin{pmatrix} 0 & \\ & 1 \end{pmatrix}_A\right] \\
		&\otimes \sqrt{p_z} \begin{pmatrix} 0 & & \\ & 1 & \\& & 1 \end{pmatrix}_B \otimes \begin{pmatrix} 1 \\ 0 \end{pmatrix}_{\tilde{A}}, \\
		K_X &= \left[ \begin{pmatrix} 1 \\ 0 \end{pmatrix}_Z  \otimes  \sqrt{\frac{p_x}{2} } \begin{pmatrix} 1 & 1 \\ 1 & 1 \end{pmatrix}_A + \begin{pmatrix} 0 \\ 1 \end{pmatrix}_Z \otimes\sqrt{ \frac{p_x}{2} } \begin{pmatrix} 1 & -1 \\ -1 & 1 \end{pmatrix}_A \right] \\
		&\otimes \sqrt{p_x} \begin{pmatrix} 0 & & \\ & 1 & \\& & 1 \end{pmatrix}_B  \otimes \begin{pmatrix} 0 \\ 1 \end{pmatrix}_{\tilde{A}}, 
	\end{aligned}
\end{equation}
 and 
\begin{equation}
	\begin{aligned}
		Z_1 &= \begin{pmatrix} 1 & \\ & 0 \end{pmatrix} \otimes \mathbb{I}_{\dim_A \times\dim_B \times 2}, \\
		Z_2 &= \begin{pmatrix} 0 & \\ & 1 \end{pmatrix} \otimes \mathbb{I}_{\dim_A \times\dim_B \times 2}.
	\end{aligned}
\end{equation}
The operators for $F^\prime$ can be constructed from Eq.~\eqref{eq:new_generic_newkraus},

\section{Bell-diagonal States} \label{appendix:bell}
For Bell-diagonal states, we can show that the announcement of the location of errors $W$ leaks no new information to Eve, by showing that Eve's state is block-diagonal in $W$ anyway.
In the Bell-diagonal case, the state shared between Alice and Bob can be written as 
	\begin{equation}
		\begin{aligned}
		\rho_{AB}&=\lambda_0 \ket{\phi_+} \bra{\phi_+} + \lambda_1 \ket{\phi_-}\bra{\phi_-}\\
		&+ \lambda_2 \ket{\psi_+}\bra{\psi_+}+\lambda_3 \ket{\psi_-}\bra{\psi_-},
		\end{aligned}
	\end{equation}
	where $\ket{\phi_{+/-}}, \ket{\psi_{+/-}}$ are the Bell states, and $\lambda_i$s are related to quantum bit error rate (QBER) via, $Q_Z=\lambda_3+\lambda_4$, $Q_X=\lambda_2+\lambda_4$, $Q_Y=\lambda_2+\lambda_3$. We can assume Eve holds a purification of the form 
	\begin{equation}
		\begin{aligned}
		\ket{\psi}_{ABE}&=\sqrt{\lambda_0} \ket{\phi_+} \ket{e_0} + \sqrt{\lambda_1} \ket{\phi_-} \ket{e_1} \\
		&+ \sqrt{\lambda_2} \ket{\psi_+} \ket{e_2}+\sqrt{\lambda_3 }\ket{\psi_-} \ket{e_3},
		\end{aligned}
	\end{equation}
	where $\ket{e_i}$ are orthonormal basis vectors for Eve's system.
Let us suppose Alice and Bob measure in the basis $\alpha\in \{X,Z\}$, the (unnormalized) state after the measurement is given by
	\begin{equation}
		\rho^{(\alpha)}_{XYE}=\sum_{x,y \in\{0,1\}}  \ket{x} \bra{x} \otimes \ket{y} \bra{y} \otimes \rho^{(\alpha),x,y}_E ,
	\end{equation}
	where $\rho^{(\alpha),x,y}_E=\operatorname{Tr} [(P^A_{(\alpha,x)} \otimes P^B_{(\alpha,y)} \otimes I_E ) \ket{\psi}\bra{\psi}_{ABE}]$. 
A simply calculation shows that the support $(\rho^{(\alpha),0,0}_E, \rho^{(\alpha),1,1}_E)$ is orthogonal to the support of  $(\rho^{(\alpha),1,0}_E, \rho^{(\alpha),0,1}_E)$. Thus, we can conclude that Eve can be assumed to always know the value of $x \oplus y$ for the entire raw key, if the state shared between Alice and Bob is Bell-diagonal. In fact, the above discussion is also true when Alice and Bob measure in the $Y$ basis, and is therefore also applicable to the six-state protocol.

\section{Twirling reduces key rate}
\label{appendix:twirling}
For our protocol, $f(\rho) = S(Z|E \tilde{A} \tilde{B})_{\rho}$. One can always expand
\begin{equation} \label{eq:S_expansion1}
	\begin{aligned}
		S(Z|E \tilde{A} \tilde{B}) &=  \sum_{\alpha,\beta} \text{Prob}(\alpha,\beta) S(Z|E,\tilde{A}=\alpha, \tilde{B}=\beta) \\
		& = \sum_\alpha  \text{Prob}(\alpha)  S(Z|E,\tilde{A}=\alpha),
	\end{aligned}
\end{equation}
where we used the fact that $Z=\bot$ for basis mismatch, and those signals are thrown away. Now, let
\begin{equation}
	\mathcal{T}(\rho) = \frac{1}{4} \sum_{i=1}^4 \rho_i = \frac{1}{4} \sum_{i=1}^4 (\sigma_i \otimes \sigma_i) \rho (\sigma_i \otimes \sigma_i)^\dagger .
\end{equation} 
Then, 
\begin{equation} \label{eq:S_expansion2}
	\begin{aligned}
 S(Z|E,\tilde{A}=\alpha)_{ \mathcal{T} (\rho)} &\leq \frac{1}{4} \sum_{i=1}^4 S(Z|E,\tilde{A}=\alpha)_{ \rho_i} \\
 &= \frac{1}{4} \sum_{i=1}^4 S(Z|E,\tilde{A}=\alpha)_{\rho} \\
 & = S(Z|E,\tilde{A}=\alpha)_{\rho},
 \end{aligned}
\end{equation}
where we have used linearity of $\mathcal{T}$ and concavity of conditional entropy in the first inequality. The second line follows from the fact that the action of the pauli operators on $\rho$ either leave the measurements performed ($X,Y,$ or $Z$) unchanged, or flip the outcomes, neither of which can affect the entropy.
Combining Eqs.~\eqref{eq:S_expansion1},\eqref{eq:S_expansion2}, we obtain  Eq.~\eqref{eq:first_condition}, which is required for the reduction to Bell-diagonal states.

\section{Decoy Analysis}
\label{appendix:decoy}
The decoy analysis in this work is similar to that from \cite{wang2021Numerical}, with small changes in notation, and is included here for sake of completeness. For a phase-randomized weak coherent pulse (WCP), the state is diagonal in photon number and follows the poissonian probability distribution 
	\begin{equation}
		p_{\mu_i} (n) = \frac{\mu_i^n}{n !} e^{-\mu_i}.
	\end{equation}
For any statistic $\gamma_{y|x}$, one can then write 
	\begin{equation}
		\gamma^{\mu_i}_{y|x} = \sum _n	p_{\mu_i} (n) \gamma^n_{y|x} ,
	\end{equation}
	where $\gamma^{\mu_i}_{y|x}$ denotes the probability of Bob obtaining outcome $y$ given Alice sent signal $x$ and intensity $\mu_i$.  
 If one uses multiple intensities, then one can use the following set of equations 
	\begin{equation}
		\begin{aligned}
			\gamma^{\mu_i}_{y|x} &\leq \sum_{n \leq N} p_{\mu_i}(n) \gamma^n_{y|x} + (1- \sum_{n \leq N} p_{\mu_i}(n) ), \\
			\gamma^{\mu_i}_{y|x} &\geq \sum_{n \leq N} p_{\mu_i}(n) \gamma^n_{y|x},
		\end{aligned}
	\end{equation}
	to obtain upper bounds  and lower bound  on $\gamma^1_{y|x}$. 
	\begin{equation}
		\gamma^{1,L}_{y | x} \leq \gamma^1_{y|x} \leq  \gamma^{1,U}_{y|x}, \quad \forall x,y 
	\end{equation}

		Noting that we can now compute bounds on $\gamma^{1}_{x,y} = \Pr(x) \gamma^{1}_{y|x}$, we obtain bounds on the all single-photon statistics for any particular coarse-graining, which we refer to as 
	\begin{equation}
		\Sfeas_1^\prime  = \{ \rho \in H_+ | \gamma^{1,L}_k \leq  \text{Tr}(\Gamma_k \rho)    \leq \gamma^{1,U}_k, \forall k\}
	\end{equation}

	where $\gamma^{1}_k$ means the $k$th statistics obtained from $1$ photon signals, and the range of $k$ depends on the exact nature of the coarse-graining.

	\textbf{Objective function:} The state shared between Alice and Bob after source-replacement can be assumed to be block-diagonal in the photon number of Alice's signal, given by
	\begin{equation}
		\rho_{AA_SB} = \sum_n p_n \ket{n}\bra{n}_{A_S} \otimes \rho^{(n)}_{AB},
	\end{equation} where $A$ and $B$ are Alice and Bob's systems, and $A_S$ is a shield system. 
	In such cases, the objective function can be shown to satisfy \cite{li2020Improving}
	\begin{equation}
		\min_{\rho \in \Sfeas} f(\rho) = \sum_n p_n \min_{\rho^{(n)}_{AB} \in \Sfeas^\prime_n} f(\rho^{(n)}_{AB}).
	\end{equation}
	For polarization encoded phase-randomized pulses, Eve can perform a photon-number-splitting attack \cite{lutkenhaus2002Quantum}. This implies that no key can be generated for $n > 1$ in the above equation. Therefore, we have
	\begin{equation}
		\begin{aligned} \label{eq:objfunc_diffn}
			\min_{\rho \in \Sfeas} f(\rho) &=  p_0 \min_{\rho^{(0)}_{AB} \in \Sfeas^\prime_0} f(\rho^{(0)}_{AB}) + p_1 \min_{\rho^{(1)}_{AB} \in \Sfeas^\prime_1} f(\rho^{(1)}_{AB}) \\
			&\geq p_1 \min_{\rho^{(1)}_{AB} \in \Sfeas^\prime_1} f(\rho^{(1)}_{AB}).
		\end{aligned}
	\end{equation}
	In this work, we will use the second expression above. 
	Since one does not know the exact single-photon statistics, but rather knows bounds on them due to decoy analysis, the optimization problem is then given by
	\begin{equation}
		\begin{aligned}
			F &= \min_{\rho \in \Sfeas_1^\prime (G)} f(\rho), \\
				\Sfeas_1^\prime & = \{ \rho \in H_+ | \gamma^{1,L}_k \leq  \text{Tr}(\Gamma_k \rho)    \leq \gamma^{1,U}_k, \forall k\}
		\end{aligned}
	\end{equation}

\end{document}